\documentclass[12pt]{article}

\advance\topmargin by -1.250in
\usepackage{booktabs}
\usepackage{multirow}
\usepackage{times}
\usepackage{mathptm}
\usepackage{graphicx}
\usepackage{array}
\usepackage{url}
\usepackage{color}
\usepackage{algorithm}
\usepackage{algorithmic}
\usepackage{amsmath}

\begin{document}

\title{Mirrored and Hybrid Disk Arrays: \\
Organization, Scheduling, Reliability, and Performance}

\author{
Alexander Thomasian
\footnote{Thomasian \& Associates, 17 Meadowbrook Road, Pleasantville, NY 10570, USA, alexthomasian@gmail.com}
}                                
\date{}
\maketitle

\vspace{-5mm}
\begin{abstract}
Basic mirroring (BM) classified as RAID level 1 replicates data on two disks,
thus doubling disk access bandwidth for read requests.
RAID1/0 is an array of BM pairs with balanced loads due to striping.
When a disk fails the read load on its pair is doubled,
which results in halving the maximum attainable bandwidth.
We review RAID1 organizations which attain a balanced load upon disk failure,
but as shown by reliability analysis tend to be less reliable than RAID1/0.
Hybrid disk arrays which store XORed instead of replicated data
tend to have a higher reliability than mirrored disks,
but incur a higher overhead in updating data.
Read request response time can be improved by processing them at a higher priority than writes,
since they have a direct effect on application response time.
Shortest seek distance and affinity based routing both shorten seek time.
Anticipatory arm placement places arms optimally to minimize the seek distance.
The analysis of RAID1 in normal, degraded, and rebuild mode is provided to quantify RAID1/0 performance.
We compare the reliability of mirrored disk organizations
against each other and hybrid disks and erasure coded disk arrays.
RAID reliabilities can be compared with a shortcut reliability analysis method.
\end{abstract}

Categories and Subject Descriptors: 
B8.1 [{\bf Performance and Reliability}]: Fault-tolerance -repllication.
C.4  [{\bf Performance of Systems}]: Reliability.
D4.2 [{\bf Operating Systems}]: Storage Management - Secondary Storage.

General terms. RAID, mirrored disks, disk scheduling, data layout.

Additional Key Words and Phrases: Reliability models, queueing theory.


\vspace{-7mm}
\section{Introduction}\label{sec:intro}

The original {\it Redundant Arrays of Inexpensive/Independent Disks (RAID)} paradigm 
was based on replacing expensive 3390 large-form factor, 
high capacity disks used by IBM mainframes with inexpensive small form factor, 
small capacity, commodity {\it Hard Disk Drives - HDDs} used by personal computers \cite{PaGK88}. 
{\it Hard Disk Drives - HDDs} are reviewed in Appendix I.
The {\it Disk Array Controller - DAC} emulates {\it Extended Count Key Data - ECKD} 
with variable block sizes on disks with fixed sized 512 byte sectors.
Since large for factor disks are not manufactured anymore,
the term Inexpensive was replaced with Independent in the RAID abbreviation \cite{Che+94}.

Fault-tolerance in the form of replication and erasure coding 
was introduced to deal with the lowered reliability resulting 
from the large number of less reliable disks required to replace a large disk. 
RAID5 is the simplest form of erasure coding with the capacity 
of a single disk out of $N$ dedicated to parity,
while RAID6 utilizes the capacity of two disks for this purpose \cite{Che+94}.
Erasure coding in RAID is discussed in \cite{ThBl09,Thom14} and reviewed in Appendix II. 

This review paper is mainly concerned with replication based on mirroring or shadowing and its variations.
Mirroring classified as RAID1 was used in early high performance systems, 
such as Tandem's NonStop SQL \cite{Tand87} and Teradata DBC/1012 computer \cite{Tera85}.
Tandem used {\it Basic Mirroring - BM}: two disks cross-connected to two processors 
to tolerate disk as well as processor failures.
Scalability was attained by two levels of high bandwidth busses.
The Teradata DBC/1012 database computer used {\it Interleaved Declustering (ID)},
so that the data on each disk is replicated on the remaining disks in the cluster
resulting in a lower increase in disk loads than BM if a disk fails \cite{Tera85}.
EMC's Symmetrix was an early successful RAID product based on mirroring.
\footnote{\url{https://en.wikipedia.org/wiki/EMC_Symmetrix}.}

Another feature of RAID is striping to balance disk loads 
by partitioning large files into fixed size stripe units or strips, 
which are placed round-robin across disk rows or stripes.
Striping is the main feature of RAID0 which has no redundancy,
but is implemented in almost all RAID arrays.

While erasure coding is much more efficient from the viewpoint of space efficiency,
{\it Hadoop Distributed File System - HDFS} uses three-way replication,
which provides reliability and parallel access provided by HDD storage \cite{CRT+13}.
HDFS places one replica on the local node, another replica on a different node at the local rack,
and the last replica on different node at a different rack. 
This policy improves write performance while not impacting data reliability or read performance.

{\it Solid State Disks (SSDs)} in the form of Flash memory provide short access time, 
consumes less power than magnetic HDDs,
are less costly than {\it Dynamic Random Access Memory (DRAM)} per byte and are non-volatile.
Flash memories can sustain a finite number of program/erase cycles,
i.e., there is wearout due to repeated writing to the same location
and writing is only possible to areas which have been pre-erased.
Specialized erasure codes have been developed to deal with localized errors in SSDs \cite{Thom14a}. 

HDDs remain the current workhorse for data storage 
and are considered in our discussion of mirroring for three reasons:
(1) most work on mirroring has been done in the context of HDDs.
(2) mirroring is too expensive to implement with flash memories.
(3) 380K PetaBytes (PB) ($10^{15}$ bytes) of HDDs 
were shipped versus 28K PB of NAND Flash in 2012 \cite{FoDH13}.

We survey research on mirrored and hybrid disk arrays, 
which instead of storing replicated blocks store XORed blocks.
We review the more influential papers, 
scattered over a large number of publications since 1990.
Video-on-demand and multimedia server data layouts are beyond the scope of this survey.
Most papers covered in this survey precede the USENIX conference on 
{\it File and Storage Technologies - FAST} since 2002,
and ACM {\it Transactions on Storage - TOS} since 2005,
which publishes independent papers and selected papers from FAST 
and the {\it Symposium on Mass Storage System Technologies - MSST} since 1974.

The paper is organized as follows: 
Appendix I provides a description of magnetic HDDs and 
their organization, disk scheduling, and data placement \cite{JaNW08}.
Section~\ref{sec:org} describes several mirrored and hybrid RAID organizations, 
where the latter maintain a 50\% redundancy level, while storing XORs of multiple data blocks. 
Performance improvement by judicious routing of disk requests is discussed in Section \ref{sec:routing}.  
Section~\ref{sec:NVRAM} discusses efficient processing of disk writes in RAID1. 
{\it Non-Volatile RAM/Storage -NVRAM/NVS} allows the caching of dirty blocks, 
so that the their destaging can be deferred and while one disk is being written, the other disk is being read. 
Studies of multi-arm disks and disks with more than one Read/Write (R/W) head on a single arm, 
to achieve lower seek distances are discussed in Section~\ref{sec:multiarm}. 
Section~\ref{sec:seek} discusses analytic models to estimate seek distances
and also cylinder remapping to reorganize data in mirrored disks to reduce seek distances.
Performance of mirrored and hybrid arrays in normal, degraded, 
and rebuild mode is discussed in Section~\ref{sec:perf} 
In Section~\ref{sec:rel} we compare the reliability of mirrored disks against each other and hybrid arrays, 
but also against erasure coded RAID arrays reviewed in Appendix II.
In Section~\ref{sec:hetero} we discuss arrays which combine multiple RAID levels. 
We conclude with Section~\ref{sec:conclusions}
Commonly used abbreviations are given preceding Appendix I.

\vspace{-5mm}
\section{Mirrored and Hybrid Disk Organizations}\label{sec:org}

There have been numerous proposals for RAID1 organizations, 
which provide a more balanced disk load when a disk fails,
but a more important consideration is RAID1 reliability as discussed in Section~\ref{sec:rel}.
{\it Hybrid Disk Arrays - HDAs} store redundant data in the form of {\it eXclusive-ORed - XORed}  data blocks.
HDAs are more reliable than RAID1,
 but incur more disk accesses for updating data.

\vspace{-2mm}
\subsection{Basic Mirroring (BM)} 

{\it Basic Mirroring (BM)} is the original form of disk mirroring, 
which replicates data on two identical disks, 
but data can also be replicated on a storage medium with equal capacity.
Similarly to other RAID1 organizations BM has the advantage 
of doubling disk access bandwidth for read requests,
but if a disk fails the read load of the surviving disk is doubled.
This is especially a problem if the data is not striped and the mirrored pair is heavily loaded.
Most performance studies of RAID1 have been conducted in the context of the BM organization.

For higher volumes of data there are $M$ pairs of mirrored disks with BM organization,
so that the total number of disks is $N=2M$.
RAID1/0 is a hierarchal RAID organization with mirrored pairs at the lower level,
which may be considered a single virtual disk with higher reliability, and RAID0 at the higher level.
Up to $M$ disk disk failures can be tolerated, as long as there are no pairs.
so that the probability of data loss due to a second disk failure is $1/M$.

RAID0/1 mirrors two RAID0 arrays, each with $M$ disks as shown in Figure~\ref{fig:BM}.
Each RAID0 arrays is considered a superdisk, 
which is considered failed when a single disk fails, 
so that up to $M$ disk failures can be tolerated as long as they are all on one side.
The probability of data loss due to a second disk failure is: $M/(2M-1)>0.5$,
which is much higher than the same probability for RAID1/0.

RAID1/0 and RAID0/1 are examples of nested RAID levels,
\footnote{\url{https://en.wikipedia.org/wiki/Nested_RAID_levels}.}
whose reliability was analyzed in \cite{Thom06b} and in Section~\ref{sec:rel}.
{\it Hierarchic RAID - HRAID} uses RAID5 erasure coding as both levels \cite{ThTH12}.

\vspace{-2mm}
\subsection{Group Rotate Declustering - GRD}\label{sec:GRD} 

GRD is a RAID0/1 array with mirroring at the higher level
with $M$ primary disks on one side,
while the $M$ secondary data at the other side are rotated 
from row to row as shown in Figure~\ref{fig:GRD}. 
GRD has the advantage that upon the failure of a disk on either side
its read accesses are evenly distributed over the $M$ disks at the other side.
GRD can tolerate up to $M$ disk failures on one side,
but fails if a second disk fails at the other side,
so that the probability of data loss similarly to RAID0/1 is $M/(2M-1) > 0.5$,
The fraction of GRD requests routed can be adjusted to balance disk loads as discussed in Section~\ref{sec:perf}.

Two variants of data placements for GRD are shown in Figure 3 in \cite{ChTo96}.                              
Disk space may be split by allocating primary data on outer cylinders and secondary data on inner cylinders.
Disk capacity is split into halfs by allocating half of outer (resp. inner) cylinders 
to primary (resp. secondary) data.
In zoned disks tracks on outer cylinders hold more sectors than inner tracks in zoned disks \cite{JaNW08},
so that primary data will occupy fewer disk cylinders than secondary data,
which implies shorter seeks in accessing randomly placed primary blocks. 
Primary (resp. secondary) data can be allocated at upper (resp. lower) tracks of a disk cylinder, 
but this is restricted to disks with an even number of tracks per cylinder.

RAID-X proposed in \cite{HwJH02} has a similarity to GRD as shown in Figure \ref{fig:RAIDx}.
Data strips on primary disks are placed diagonally in secondary areas.

\begin{figure}
\renewcommand\arraystretch{1.2}
\centering
 \begin{tabular}{|c c c c||c c c c|}
 \hline
 \multicolumn{4}{|c||}{Primary Disks} &  \multicolumn{4}{c|}{Secondary Disks}\\
 \hline
 $D_1$ & $D_2$ & $D_3$ & $D_4$ & $D_5$ & $D_6$ & $D_7$ & $D_8$ \\
\hline
 $A$ & $B$ & $C$ & $D$ & $A'$ & $B'$ & $C'$ & $D'$ \\
 \hline
 $E$ & $F$ & $G$ & $H$ & $E'$ & $F'$ & $G'$ & $H'$ \\
 \hline
 $I$ & $J$ & $K$ & $L$ & $I'$ & $J'$ & $K'$ & $L'$ \\
 \hline
 $M$ & $N$ & $O$ & $P$ & $M'$ & $N'$ & $O'$ & $P'$ \\
\hline
 \end{tabular}
\caption{\label{fig:BM}RAID0/1 with $N=8$ disks. 
Rotated primed blocks replicate unprimed blocks.}

\hspace{1mm}

\renewcommand\arraystretch{1.2}
\centering
 \begin{tabular}{|c c c c||c c c c|}
 \hline
 \multicolumn{4}{|c||}{Primary Disks} &  \multicolumn{4}{c|}{Secondary Disks}\\
 \hline
 $D_1$ & $D_2$ & $D_3$ & $D_4$ & $D_5$ & $D_6$ & $D_7$ & $D_8$ \\
\hline
 $A$ & $B$ & $C$ & $D$ &   $A'$ & $B'$ & $C'$ & $D'$ \\
 \hline
 $E$ & $F$ & $G$ & $H$ &   $H'$ & $E'$ & $F'$ & $G'$ \\
 \hline
 $I$ & $J$ & $K$ & $L$ &   $K'$ & $L'$ & $I'$ & $J'$ \\
 \hline
  $M$ & $N$ & $O$ & $P$ &  $P'$ & $M'$ & $N'$ & $O'$  \\
\hline
 \end{tabular}
\caption{\label{fig:GRD}Group Rotate Declustering with $N=8$ disks.}

\hspace{1mm}

\renewcommand\arraystretch{1.2}
\centering
 \begin{tabular}{|c c c c|}  
 \hline
 \multicolumn{4}{|c||}{Cluster 1} \\ 
 \hline
 $D_1$ & $D_2$ & $D_3$ & $D_4$  \\ 
 \hline
 $~B_0~$ & $~B_1~~$ & $~B_2~~$ & $~B_3~$  \\
 \hline
 $~B_4~$ & $~B_5~~$ & $~B_6~~$ & $~B_7~$  \\
 \hline
 $~B_8~$ & $~B_9~~$ & $~B_{10}~~$ & $~B_{11}~$  \\
 \hline 
\hline
 $~M_9~$  & $~M_6~~$ & $~M_3~~$ & $~M_0~~$  \\
\hline
 $~M_{10}~$ & $~M_7~~$ & $~M_4~~$ & $~M_1~~$  \\
\hline
 $~M_{11}~$ & $~M_8~~$ & $~M_5~~$ & $~M_2~~$  \\
\hline
 \end{tabular}
\caption{\label{fig:RAIDx} RAID-x architecture with $M=4$ disks in one cluster.}

\hspace{1mm}

\renewcommand\arraystretch{1.2}
\centering
 \begin{tabular}{|c c c c||c c c c|}
 \hline
 \multicolumn{4}{|c||}{Cluster 1} &  \multicolumn{4}{c|}{Cluster 2}\\
 \hline
 $D_1$ & $D_2$ & $D_3$ & $D_4$ & $D_5$ & $D_6$ & $D_7$ & $D_8$ \\
\hline
\hline
 $~A~$ & $~B~~$ & $~C~~$ & $~D~~$ & $~E~~$ & $~F~~$ & $~G~~$ & $~H~~$ \\
 \hline
 $b_3$ & $a_1$ & $a_2$ & $a_3$ & $f_3$ & $e_1$ & $e_2$ & $e_3$ \\
 \hline
 $c_2$ & $c_3$ & $b_1$ & $b_2$ & $g_2$ & $g_3$ & $f_1$ & $f_2$ \\
 \hline
 $d_1$ & $d_2$ & $d_3$ & $c_1$ & $h_1$ & $h_2$ & $h_3$ & $g_1$ \\
\hline
 \end{tabular}
\caption{\label{fig:ID} Interleaved Declustering with $N=8$ disks, $c=2$ clusters, and $n=4$ disks per cluster.
Capital letters denote primary data and small letters subsets of secondary data.}

\hspace{1mm}

\renewcommand\arraystretch{1.2}
\centering
 \begin{tabular}{|c c c c c c c c|}
 \hline
 $D_1$ & $D_2$ & $D_3$ & $D_4$ & $D_5$ & $D_6$ & $D_7$ & $D_8$ \\
\hline
 $A$ & $B$ & $C$ & $D$ & $E$ & $F$ & $G$ & $H$ \\
 \hline
 $h$ & $a$ & $b$ & $c$ & $d$ & $e$ & $f$ & $g$ \\
\hline
 \end{tabular}
\caption{\label{fig:CD} Chained declustering with $N=8$ disks. 
Primary blocks are in caps and secondaries in small letter.}
\end{figure}

\vspace{-2mm}
\subsection{Interleaved Declustering (ID)}\label{sec:ID} 

ID organizes $N$ disks into $c$ clusters with $n=N/c$ disks per cluster,
e.g., $N=8$, $c=2$, and $n=4$ as shown in Figure~\ref{fig:ID}.
Disk data placement may follow GRD, 
with half of disk capacity dedicated to primary data and the other half to secondary data.
The primary data on each disk is partitioned into $n-1$ equal size areas storing secondary data. 

The maximum number of disk failures that can be tolerated by ID is $I=c$,  
while $I=M$ for BM, GRD, and CD organizations.
The advantage of ID over BM is that the read load of primary data 
on a failed disk is distributed over $n-1$ disks.

The {\it Striped Mirroring Disk Array (SMDA)} described in \cite{JiHw00} has a layout similar to ID.

\vspace{-2mm}
\subsection{Chained Declustering (CD)}\label{sec:CD} 

The CD organization was proposed in conjunction with the Gamma database machine project \cite{HsDe90},
which was inspired by the Teradata DBC/1012 database computer \cite{Tera85}.
Disk space is shared equally between primary and secondary areas,
so that primary data on disk $D_i$ is replicated 
as secondary data on disk $D_{i+1\mbox{mod}(N)}$ as shown in Figure~\ref{fig:CD}.
CD can tolerate up to $M$ disk failures, as long as there are no consecutive disks.

Data can be accessed from its primary or secondary areas, similarly to ID and GRD organizations.
Unlike ID, CD does does spread secondary data over multiple disks,
but CD can be extended by allowing the primary data to be distributed over the next $n-1$ disks.
While such a CD array results in a more balanced load, it is less reliable, 
since data loss will occur if the two failed disks are within $n$ of each other.

A shown in Figure~\ref{fig:CD'} in the case of disk failures fractional routing 
of read requests can attain a balanced read load \cite{HsDe90,ThXu08}. 
With two disk failures the routing probabilities should be adjusted,
as shown by the last two rows of Figure~\ref{fig:CD'}.

\begin{figure}
\renewcommand\arraystretch{1.2}
\centering
 \begin{tabular}{|c c c c c c c c|}
 \hline
 $D_1$ & $D_2$ & $D_3$ & $D_4$ & $D_5$ & $D_6$ & $D_7$ & $D_8$ \\
\hline
\hline
 $1 \times A$ & $0 \times B$ & $\frac{1}{7}C$ & $\frac{2}{7}D$ & $\frac{3}{7}E$ & $\frac{4}{7}F$ & $\frac{6}{7}G$ & $\frac{6}{7}H$ \\
 \hline
 $\frac{1}{7}h$ & $0 \times a$ & $b$ & $\frac{6}{7}c$ & $\frac{5}{7}d$ & $\frac{4}{7}e$ & $\frac{3}{7}f$ & $\frac{2}{7} g$ \\
\hline
\hline
 $1\times A$ & $0  \times B$ & $1 \times C$ & $0 \times D$ & $\frac{1}{5} E$ & $\frac{2}{5}F$ & $\frac{3}{5}G$ & $\frac{4}{5}H$ \\
 \hline
 $\frac{1}{5}h$ & $0 \times a$ & $b$ & $0 \times c$ & $1 \times d$ & $\frac{4}{5}e$ & $\frac{3}{5}f$ & $\frac{2}{5} g$ \\
\hline
 \end{tabular}
\caption{\label{fig:CD'} 
Balancing loads in CD after $D_2$ fails and $D_2$ and $D_4$ failures by routing read requests.}
\end{figure}

The Petal disk array utilizes striping (RAID0), 
but also the CD organization as shown in Figure 5 in \cite{LeTh96},
which shows that alternate rows are dedicated to primary and secondary data.

\vspace{-2mm}
\subsection{Dual or Hybrid Striping} 

Dual striping combines large and small stripes to improve performance \cite{MeYu95}. 
Database queries requiring table scans and OLTP workloads accessing small data blocks,
access data from the large and small stripes, respectively.
Having data distributed over small stripes reduces data access skew for OLTP applications,
while fewer seeks are required when table scans are processed over large stripes.
Note that the ID organization implicitly utilizes two strip sizes,
where secondary strips are $n-1$ times smaller than primary strips.

Disks are modeled as M/G/1 queues \cite{Klei75}, which are briefly described in Section \ref{sec:queueing}.
Disks process accesses to short blocks with Poisson arrivals,
but also one large disk access one at a time according to the {\it Permanent Customer Model (PCM)} \cite{BoCo91},
which is discussed in more detail in Section~\ref{sec:rebuild}.
The performance study showed that dual striping outperforms uniform striping 
in both normal and degraded modes (with one disk failure) for the disk workload under consideration.
Table scans may result in unacceptable delay in the processing of accesses to short blocks by OLTP transactions. 
A solution is to temporarily pause lengthy table scans 
to allow short disk accesses to proceed, as discussed in \cite{Tand87}.
This requires an extension of \cite{BoCo91} to allow the preemptive policy for permanent customers. 



\vspace{-2mm}
\subsection{LSI Logic RAID} 

LSI Logic RAID combines mirroring and parity by placing Parity disks (Pdisks) 
between pairs of Data disks (Ddisks) \cite{Wiln00}.
The {\it Data out-Degree (DoutD)} is the number of parities associated with each data element 
and the {\it Parity in-Degree (PinD)} is the number of data elements XORed to compute a parity, 
so that DoutD=PinD=2 in this case.
With $N=2M=8$ disks there are four Ddisks and four Pdisks denoted by D and P, respectively,
so that $P_{i,i+1(modM)} = D_i \oplus D_{i+1(modM)}, 1 \leq i \leq 4$, as shown below:

{\large \bf ($D_1, P_{1,2}, D_2, P_{2,3}, D_3, P_{3,4}, D_4, P_{4,1}$)}.

LSI RAID can tolerate two disk failures and 
three consecutive disk failures in one half of cases,
when the middle disk is a Pdisk, but not otherwise.
If the three disks ($D_1,P_{1,2},D_2$) fail then recovery can proceed as follows:
$D_1 = D_4 \oplus P_{4,1}$, $D_2 = D_3 \oplus P_{2,3}$, and $P_{1,2} = D_1 \oplus D_2$.

An OLTP workload with read and write accesses 
to small data blocks will result in an unbalanced disk load if all updates, 
which are uniform over data blocks are carried out as {\it Read-Modify-Wrote (RMW)} accesses.
This load is balanced in \cite{ThTa12} by using a combination of RMW 
and {\it ReConstruct Write (RCW)} accesses \cite{Thom05a}.
Disk loads can also be balanced as shown below by shifting the strips right from row-to-row, 
where $d$s and $p$s represent strips in the first row, 
and $\bar{d}$ and $\bar{p}$ represent the shifted strips in the second row. 
The disadvantage of this layout is that it does not tolerate consecutive three disk failures.

\noindent
{\large \bf ($d_1, p_{1,2}, d_2, p_{2,3}, d_3, p_{3,4}, d_4, p_{4,1}$)}.  \newline
{\large \bf ($\bar{p}_{4,1}, \bar{d}_1, \bar{p}_{1,2}, \bar{d}_2, 
\bar{p}_{2,3}, \bar{d}_3, \bar{p}_{3,4}, \bar{d}_4 $)}.

\vspace{-2mm}
\subsection{SSPiRAL (Survivable Storage using Parity in Redundant Array Layout)}\label{sec:SSP} 

Several mirrored and hybrid disk arrays (HDAs) are investigated in \cite{ALPS08}.
There are $N=2M$ disks where $M$ disks are Ddisks and the other $M$ are Pdisks.
LSI Logic RAID is SSPiRAL(4+4,2) with four Ddisks and four Pdisks and DoutD=PinD$=2$, 
as shown in Figure~\ref{fig:SSP822}.

\begin{figure}[htb]
\renewcommand\arraystretch{1.2}
\centering
 \begin{tabular}{|c c c c c c c c|}
\hline
  $D_1$ & $D_2$ & $D_3$ & $D_4$ & $D_5$ & $D_6$ & $D_7$ & $D_8$  \\
\hline
 $d_1$ & $p_{1,2}$ & $d_2$ & $p_{2,3}$ & $d_3$ & $p_{3,4}$ & $d_4$  & $p_{4.1}$ \\
\hline
 \end{tabular}
\caption{\label{fig:SSP822} SSPiRAL(4+4,2) with four Ddisks and four Pdisks with PinD=DoutD=2.}
\end{figure}

SSPiRAL extends LSI RAID to DoutD=PinD=3, so that each Ddisk is protected by three Pdisks and vice-versa.
SSPiRAL(4+4,3) with $N=8$ disks, four Ddisks and four Pdisks is shown in Figure~\ref{fig:SSP443}.

\begin{figure}[htb]
\renewcommand\arraystretch{1.2}
\centering
 \begin{tabular}{| c  c  c  c  c  c  c  c  |}
\hline
 $D_1$ & $D_2$ & $D_3$ & $D_4$ & $D_5$ &  $D_6$ &  $D_7$  & $D_8$     \\
\hline
$d_1$ & $d_2$ & $d_3$ & $d_4$ & 
{ $d_1 \oplus d_2 \oplus d_3$ }  &
{ $d_2 \oplus d_3 \oplus d_4$ }  &
{ $d_3 \oplus d_4 \oplus d_1$ }  &
{ $d_1 \oplus d_2 \oplus d_3$ }  \\
\hline
\end{tabular}
\caption{\label{fig:SSP443} SSPiRAL(4+4,3) with PinD=DoutD=3.}
\end{figure}

For SSPiRAL(4+4,3) all disk failures up to three 
can be tolerated $A(N,i)= {N \choose i}, 0 \leq i \leq 3$.
There are ${8 \choose 4}=70$ possibilities for four disk failures and data loss occurs in 14 cases \cite{ALPS08}, as follows:
(1) A Ddisk and the 3 Pdisks in which it participates (4 cases).
(2) Two out of four Ddisks and the two Pdisks in which they both participate (6 cases).
(3) Three out of four Ddisks and the Pdisk in which all three participate (4 cases).
If all four Ddisks fail they can be reconstructed using Pdisks,
e.g., $\{ d_1 \oplus d_2 \} = \{ d_3 \oplus d_4 \oplus d_1 \} \oplus \{ d_2 \oplus d_3 \oplus d_4 \}$ and 
$d_4 = \{ d_1 \oplus d_2 \} \oplus \{d_4 \oplus d_1 \oplus d_2 \}$.

\vspace{-2mm}
\subsection{B-code}\label{fig:Bcode} 

For B-code and its dual parities associated with data are stored in each column \cite{XBBW99}.
RM2 and X-code are two 2DFT arrays with horizontal parities, 
where the former is not MDS and the latter is MDS \cite{ThBl09},
but B-codes like Weaver codes, discussed below, store data on the same device \cite{Hafn05}. 
In the case of $n=3$ bits per column and $\ell= 2n =6$ columns, 
we have $\hat B_6$, which is a dual of $B_6$ code, 
as shown below based on Figure 1 in \cite{XBBW99}:

\vspace{-2mm}
\[
\left(
\begin{matrix}
a_1 & a_2 & a_3 & a_4 & a_5 & a_6  \\
a_2+a_3 & a_3+a_4 & a_4+a_5 & a_5+a_6 & a_6+a_1 & a_1+a_2 \\
a_4+a_6 & a_5+a_1 & a_6+a_2 & a_1+a_3 & a_2+a_4 & a_3+a_5 
\end{matrix}
\right)
\]
When the four columns 3-6 are erased they can be reconstructed in the following manner.
$a_3 = (a_2 + a_3) \oplus a_2$,
$a_4 = (a_3 + a_4) \oplus a_3$,
$a_5 = (a_5 + a_1) \oplus a_1$,
$a_6 = (a_4 + a_6) \oplus a_4$.

Properties of B-codes and their duals are listed in \cite{XBBW99}.
This code is MDS since by placing the 6 data bits in first two columns 
we have four parity columns and four column failures can be tolerated.

\vspace{-2mm}
\subsection{Weaver Codes}\label{sec:Weaver} 

WEAVER codes have three design features \cite{Hafn05}:
"(a) data and parity blocks on the same strip, 
(b) constrained parity in-degree, 
(c) balance and symmetry".
There are $n$ strips per stripe and each strip holds $r$ data elements and $q$ parity elements.
For Weaver$(n,k,t)$ parity elements have an in-degree $k$ and data elements have out-degree $t$.
Given $m = \mbox{gcd}(k, t)$, $r = k/m$ and $q = t/m$ are minimal. 
The efficiency of the code is $e=r/(r+q)=k/(k+t)$.

Since $k \leq t$ the maximum efficiency of this code is 50\%.
Weaver$(n,t,t)$ codes with 50\% efficiency are a special category of interest in this study.
Table 1 in \cite{Hafn05} is a partial listing of 
{\it Parity Defining Sets (DPSs)} for WEAVER(n,t,t) codes,
where the PDS is $\kappa (j)= \kappa_1 (j) +s\mbox{mod}(n)$,
e.g., the second row for $t=3$ in the table: $\kappa_1 (0)= (1,2,4)$ and $s=2$,
so that the parities corresponding to $d_j$ are at $(j+3,j+4,j+6)\mbox{mod}(n)$.
LSI Logic RAID is Weaver(n,2,2) and SSPiRAL is Weaver(n,3,3),
but unlike Weaver codes they place parities on separate devices.

\begin{figure}
\renewcommand\arraystretch{1.2}
\centering
 \begin{tabular}{|c | c | c | c |}
\hline
  $D_1$ & $D_2$ & $D_3$ & $D_4$ \\
\hline
 $d_1$ & $d_2$ & $d_3$ & $d_4$ \\
 \hline
{ $d_2 \oplus d_3$ }  & { $d_3 \oplus d_4$ } & { $d_4 \oplus d_1$ }  & { $d_1 \oplus d_2$ }  \\
\hline
 \end{tabular}
\caption{\label{fig:WW422}Data layout for a Weaver-like (4,2,2) with $N=4$ disks.}
\end{figure}

Weaver(4,2,2) can tolerate two disk failures,
e.g., if disks $D_1$ and $D_2$ fail then $d_1 = d_4 \oplus \{d_4 \oplus d_1\}$, 
$d_2 = d_1 \oplus \{ d_1 \oplus d_2 \}$,
but unlike LSI Logic RAID three disk failure cannot be tolerated.

Weaver(8,3,3) with $N=8$ disks with PoutD=PinD=$3$ as shown in Figure~\ref{fig:SSP443}.

\begin{figure}
\renewcommand\arraystretch{1.2}
\centering
\begin{footnotesize}
 \begin{tabular}{|c | c | c | c | c | c | c | c |}
\hline
    $D_1$  &  $D_2$  &  $D_3$  &  $D_4$  &  $D_5$  & $D_6$ & $D_7$ &  $D_8$  \\
\hline
    $d_1$  &  $d_2$  &  $d_3$  &  $d_4$  &  $d_5$  & $d_6$ &  $d_7$ &  $d_8$ \\
\hline
 {$d_3  \oplus d_4 \oplus d_6$} &
 {$d_4  \oplus d_5 \oplus d_7$} &
 {$d_5  \oplus d_6 \oplus d_8$} &
 {$d_6  \oplus D_7 \oplus d_1$} &
 {$d_7  \oplus d_8 \oplus d_2$} &
 {$d_8  \oplus d_1 \oplus d_3$} &
 {$d_1  \oplus d_2 \oplus d_4$} &
 {$d_2  \oplus d_3 \oplus d_5$} \\
\hline
 \end{tabular}
\end{footnotesize}
\caption{\label{fig:WW833} Weaver(8,3,3) data layout with $N=8$ disks and $t=3$.}
\end{figure}

Since a data block and associated parities for Weaver(8,3,3) 
appear at four different disks all three disk failures can be tolerated.
Consider the failure of $D_1$, $D_4$, $D_6$, so that $d_1$'s parity is only available at $D_7$.
To recover $d_1$ we need to recover $d_4$, since $d_2$ is available.
We have the following steps $(d_4 \oplus d_5 \oplus d_7) \oplus d_5 \oplus d_7 \rightarrow d_4$
and $(d_7 \oplus d_8 \oplus d_2) \oplus d_7 \oplus d_8 \rightarrow d_2$.Finally, $d_2 \oplus d_4 \oplus (d_1 \oplus d_2 \oplus d_4) \rightarrow d_1$.


Weaver(8,3,3) can tolerate four disk failures,
as long as they do not involve a data strip and its parities.
Recovery is possible when the first four disks in Figure~\ref{fig:WW833} fail, as follows:
$d_7 \oplus d_8 \oplus (d_7,d_8,d_2) \rightarrow d_2$,
$d_2 \oplus d_5 \oplus (d_2,d_3,d_5) \rightarrow d_3$,
$d_8 \oplus d_3 \oplus (d_8,d_1,d_3) \rightarrow d_1$,
$d_1 \oplus d_2 \oplus (d_1 \oplus d_2 \oplus d4) \rightarrow d_4$.
For $N=8$ disks and four disk failures there are ${8 \choose 4}=70$ possibilities
and out of these $N=8$ rotations of a data strip and its PDS cannot be tolerated.

\vspace{-3mm}
\subsection{Robust, Efficient, Scalable, Autonomous, Reliable (RESAR)}\label{sec:RESAR} 

The RESAR exabyte scale storage places disklets into two different parity groups 
with one parity disklet in each stripe \cite{Sch+16}.
In fact RESAR is RAID5/1, i.e., a mirrored systems where each side is a RAID5 with $M$ disks.
When a disk fails its mirror is accessed first and if this fails data is reconstructed using the RAID5 paradigm.
"A RESAR-based layout with 16 data disklets 
per stripe has about 50 times lower probability of suffering data loss
in the presence of a fixed number of failures than a corresponding RAID 6 organization".
The simulation to estimate this probability for 100,000 disks required 10,000 hours.

\subsection{Multiway Placement} 

Three-way versions of chained, group rotate, and standard mirroring are specified \cite{SWZG11}.
The {\bf Shifted Declustering (SD)} provides optimal parallelism, 
since the data replicas are distributed evenly.
The similarity of SD to layouts proposed in \cite{ABSC98} requires further investigation.

\vspace{-5mm}
\section{Routing Read Requests in Mirrored Disks}\label{sec:routing}

We summarize routing strategies in mirrored disks in first following \cite{ToCY90} and then \cite{Thom06a}.

{\bf 1. Single Queue (SQ)}:
Update requests can be processed only when both disks are idle,
while read requests can be processed on both disks or only the primary disk (see 1.1).

{\bf 1.1 Primary/Secondary (PSSQ)}: 
Disks are designated as primary and secondary.
(i) {\it Serial PSSQ (S-PSSQ)} allows only one request to be processed at a time.
(ii) {\bf Concurrent PSSQ (C-PSSQ)} allows concurrency between read and update requests.
A read request can start service at the primary disk.

{\bf 1.2 Equitable SQ (ESQ)}: 
Disks are treated equally.

{\bf Concurrent Read ESQ (CR-ESQ)}: 
Allows reads to be processed concurrently.
Only one update is allowed to be processed at any one time.

{\bf Concurrent Read Update ESQ (CRU-ESQ)}: 
Allow concurrency between reads and updates.
An update is still required to wait until both disks are available,
but a read request may proceed as soon as a disk becomes available.
 
{\bf Minimum Read ESQ (MR-ESQ)}: 
Similarly to S-PSSQ each request waits until the previous request is completed.
Read requests are processed at both disks, but when the first read request completes, abort the other request.
It should be noted that disk accesses cannot be preempted at all stages, especially during seeks.

{\bf 2. Multiple Queue Policies}:

{\bf Distributed MQ (DMQ)}: 
Maintain a separate queue for each disk. 
An update request generates write requests for each queue.
Read requests are routed randomly to balance disk loads. 

{\bf Shortest Queue - DMQ (SQ-DMQ)}: 
Route read requests to the shortest queue.
Reads can also spawn requests at both queues.
There are two variations Minimum Read - MR with and without preemption.
Abort is initiated when a read request begins service or is completed. 

{\bf Common MQ (CMQ)}:
The request at the head of the queue starts processing as soon as either disk is free.
Update requests are spawned at both disks and are enqueue at the busy disk.

RAID1 performance is compared with RAID5 in \cite{ChTo96}.
Three RAID1 configurations are considered BM, CD, and GRD,
where GRD is shown to provide the best performance for small and large I/O environments.
Three routing policies are considered for RAID1: 
Random Join (RJ), Shortest Queue (SQ), and Minimum Seek (MS).
Local disk scheduling policies can additionally be applied at each disk.

Routing policies described in \cite{ToCY90} are analyzed in \cite{ToCh91} 
using the {\it Matrix Geometric Method (MGM)} \cite{Neut81}.
The processing of read and write requests in mirrored disks 
have similarities to their scheduling in replicated databases, which is reviewed in \cite{Thom14b}.


Routing of disk accesses in RAID1 is classified according to queue organizations in \cite{Thom06a}:

\begin{description}

\item
[Private Queue (PQ) with Immediately Routing:]
Based on a routing criterion read requests are sent to one of the disks,
while write requests are sent to both disks.

\item
[Shared Queue (SQ) with Deferred Routing:] 
Read requests held at SQ are routed to the first disk that becomes available,
while immediate routing may result in idle disk, while the disk has a queue.
Dynamic routing based on disk state can be pursued more efficiently than PQ.

\item
[Hybrid Queue (HQ):] 
Read requests held in SQ are routed to PQs after some delay to ensure the disk queue is not empty.

\end{description}

Two examples of static routing are: 
uniform with equal probabilities and round-robin routing.
It is shown in \cite{Thom06a} and Section~\ref{sec:perf} 
that with Poisson arrivals and exponential service times
round-robin routing improves mean response time with respect to uniform routing. 

{\it Affinity Based Routing (ABR)} divides disks to inner and outer disk blocks,
which are delineated by a pivot point, which is {\it Logical Block Address - LBA} on disk \cite{ThHa05}.
Numbering LBAs from the outermost disk track,
requests with an LBA lower than the pivot point are sent 
to the disk serving reads on outer tracks and vice-versa.
With $2s$ sectors it seems that the pivot point should be at sector $s$,
but in zoned disks this would entail outer data occupying fewer cylinders 
than inner data blocks and hence shorter seek times and access times \cite{JaNW08}, but also see Appendix I.
Several criteria to determine the pivot point in zoned disks are considered in \cite{ThHa05}.

Numerical results show that pivot point selected based on balanced disk utilizations   
achieve about the same performance as the one based on minimum mean response time,
which is justified by Eq.~\ref{eq:PK}.
The "transposed mirrored organization",
which switches the data on inner and outer tracks achieves 
the best performance for heavily loaded disks. 
We also discuss an adaptive approach to implement ABR,
so that the pivot point is determined by the router based on the history of routed requests.
Sequential reading of large files via successive block accesses benefits from ABR 
since it eliminates seeks for such accesses. 
In a situation where multiple large files are being read,
the pivot point can be assigned dynamically to split disk space logically 
to minimize seek times while balancing disk loads. 
The onboard disk buffer will further improve performance if sequential prefetching is in effect.

The dynamic {\it Join the Shortest Queue (JSQ)} policy is optimal 
when the service rates are fixed or non-decreasing \cite{Whit86}.
JSQ may not be effective for disk scheduling,  
since the disk queuelength is not a good indicator 
of the remaining processing time with FCFS scheduling,
e.g., consider accesses to neighboring blocks on a track,
which can be accessed via a single I/O, an instance of proximal I/O \cite{ScSS11}. 
Instead of estimating the response time of the request to be routed with SATF scheduling,
it might be better to reduce the mean response time over all requests \cite{Thom06a}.

The router may use the LBA to send a request to the disk processing a request with the closest LBA,
e.g., the last request in the queue with FCFS scheduling.
This is an approximation to the {\t Shortest Job First - SJF},
which is the best nonpreemptive policy for single servers \cite{Klei76}. 
Given the LBA accurate timing emulators can be used to estimate disk service time.

For random disk accesses performance is mainly determined 
by the local disk scheduling policy and not the PQ (private queue) routing policy.
SATF scheduling with SQ (shared queue) applied over all replicated disks
provides more opportunities than PQ to improve performance, 
since twice as many requests are available for SATF scheduling \cite{Thom06a}
There are no controllers for mirrored disks which could serve requests from a shared queue.

The {\it Distributed Shortest-Positioning Time First (D-SPTF)} 
protocol dynamically distribute requests in decentralized storage servers,
selecting from servers which hold a replica \cite{LuGo04}. 
For 10-200 microsecond network latencies D-SPTF performs as well as a centralized system. 
D-SPTF achieves up to 65\% higher throughput than popular decentralized approaches
and adapts more cleanly to heterogeneous server capabilities.

\vspace{-5mm}
\section{Efficient Processing of Writing to Disk in Mirrored Disks}\label{sec:NVRAM}   

Disk-resident data is cached in computer systems at the main memory buffer,
at the cache at the DAC (disk array controller), and onboard disk cache.
Large cache capacities have resulted in a reduced miss rate for read requests,
so that it has been argued that most disk accesses are writes
and it is important to reduce the disk load due to writes.

Write anywhere policy on either or both mirrored disks has been used to minimize disk arm movement
to reduce the susceptibility to data loss in systems without an NVRAM cache \cite{SoOr91,SoOr93}.
A directory keeps track of blocks written anywhere and 
to allow efficient sequential accesses data is updated later on the primary disk, 


The two-phase method for mirrored disks processes read requests at one disk, 
while the other disk is destaging dirty blocks in a batch mode using a CSCAN scheduling, 
i.e., SCAN in one direction \cite{PoBD93}.
It is unlikely that equal durations for the two phases will result in a perfect overlap in read and write processing.
Refinements of this method are addressed in \cite{ThLi05}: 
(1) Eliminate forced idleness by processing write requests individually.
This can be carried out opportunistically with "freeblock scheduling",
which accomplishes useful work during disk rotation time \cite{LuSG02}.
(2) Given the considerable rotational latency, 
instead of CSCAN use SATF or destage according 
to a permutation of outstanding requests to minimize destage time,
(3) destaging of dirty blocks is deferred if the number of enqueued read requests exceeds a threshold,
whose value is determined by monitoring the system.
This provides for more opportunities for dirty blocks to be overwritten.

Part of the DAC cache is an NVRAM, 
{\it Dynamic Random Access Memory - DRAM} \cite{JaNW08} protected by {\it Uninterruptible Power Supply (UPS)}.
A duplexed NVRAM is as reliable as a magnetic disks \cite{MeCo93}, which allows a {\it fast write} capability.
The destaging/writing of modified data blocks to disk can be deferred,
allowing reads to be processed at a higher priority than writes,
which improves transaction response time in OLTP.
Overwriting of dirty blocks in NVRAM obviates unnecessary destages to disk
and deferred writing in batches in LBA order reduces positioning time.
The results of trace analysis such as \cite{TrMe95} 
was used to quantify the number overwritten dirty blocks
and the proximity of destaged requests (number on the same track) \cite{ThMe97} 
Destage is initiated when the buffer becomes full or when it is filling rapidly. 

An NVRAM is not required for deferred destaging of dirty blocks 
in OLTP systems with {\it Write Ahead Logging (WAL)} \cite{RaGe02},
but this would entail a costly recovery process if the system crashes.
Before and after images of modified data are logged.
The before images are used in undoing the updates of aborted transactions,
while after images are required to allow the NoForce policy,
i.e., a transaction may commits without forcing dirty pages to disk and such dirty pages can be overwritten.



\vspace{-5mm}
\section{Disks with Multiple Arms and Multiple R/W Heads per Arm}\label{sec:multiarm}

Disks with two {\it Read/Write (R/W)} heads on one disk arm,
which are at a fixed distance from each other are considered in \cite{CaCF84,MaKo89}.   
The {\it Nearer-Server (NS)} rule of such a system is analyzed in two cases in \cite{CaCF84}: 
(1) both heads have to be kept on the surface of the disk.
(2) this restriction is relaxed.
It was shown that an optimization with respect to $d$ yields an expected seek distance, 
which is slightly less than that of a system with two independent arms 
and a single controller which allows the movement of one arm at a time.

It is shown in \cite{MaKo89} that for a disk with $C$ cylinders 
the optimal distance is $d=C/2-1$ between the heads 
and the seek distance is reduced from $C/3$ to $C/6$.
A disk with a two heads at a fixed distance $d$ is studied 
by simulation of an actual system in \cite{PaWo81}. 

Scheduling of a disk with two arms with a single R/W head is studied in \cite{Hofr83}.
The closest head is used to serve an incoming request,
while the other head is moved to a better position in anticipation of the next request.
If the accessed cylinder is $a < C/2$ then the head moves somewhere between $a$ and $C$
and if $a > C/2$ the head is positioned between the first cylinder and $a$ 
More specifically it is shown in \cite{MaVa91} that in the first case 
the head should be positioned at $a + \lfloor (2(C-a)/3 \rfloor$ and otherwise at $\lfloor a/3 \rfloor$.
The analysis yields a seek distance $5C/36$, which was given without proof in \cite{Hofr83}.

There is a Conner patent for multiple actuator Chinook disk \cite{SBSA94}.
\footnote{\url{https://en.wikipedia.org/wiki/Conner_Peripherals#Performance_issues_and_the_Chinook_dualactuator_drive.}}
A dual actuator logging disk architecture with one head dedicated to reading 
and the other arm to logging in regions with free disk sectors is proposed in \cite{Chan03}.

The optimal placement for two-headed disk systems is the camel arrangement,
which is two consecutive organ-pipe arrangements \cite{MaKo90}.
There are $2^{N/2+1}$ optimal camel arrangements for a disk with $N = 2(2n + 1)$ cylinders.

Improving disk performance via latency reduction in the context of variable block disks 
with {\it Rotational Position Sensing (RPS)} is addressed in \cite{NgSW91}.
Three methods are proposed to improve latency, which are also applicable to non-RPS disks: 
(1) Moving the disk arm seeking to the same cylinder on both disks.
Synchronized the two disks to be half a rotation away from each other
results in a reduction of latency from $T_{rot} /3$ to $T_{rot} /4$.
(2) Two copies of the data are placed 180 degrees out of phase from each other.
This can be done by doing so on alternating tracks (even and odd) 
or using half of the capacity of each track.
(3) Dual actuators placed opposite each other.

Given a fixed capacity disk array analytical models are developed in \cite{YGC+00} 
to find the combination of striping, mirroring, and rotational data replication, 
which yields the best performance, i.e., with reduced seek times and rotational delays,
for given workload and disk characteristics. 
The effectiveness of the configuration models are tried on a prototype.
 
A taxonomy of disk parallelism is provided in \cite{SaGS08}.
In DASH, Disk stack is the number of disk platters constituting a cylinder, 
Arm assembly is the number of arms. 
Surface is the number of surfaces, e.g., two if data recorded on both sides of platters. 
and Head (number of R/W heads per arm). 
A conventional disk is then D1 A1 S1 H1.
Air turbulence affects the vibration of platters and heads, 
which makes it impossible to transfer data simultaneously from multiple tracks.
A cost benefit analysis of intradisk parallelism is reported in this paper,
which also saves disk power consumption, a topic beyond the scope of this paper.

\vspace{-5mm}
\section{Seek Distances with Mirrored Disks}\label{sec:seek} 

The analysis for the seek distance for reads with $k$-way replication is given in \cite{BiGr88}.
The analysis is based on the fact that for a disk with $C$ cylinders or tracks
the {\it probability mass function - pmf} for uniform accesses to disk cylinders is                                                                    
$P[X=i] = 2(C-i)/C^2, 1 \leq i \leq C-1$, $P[ X \geq i ] \approx (1-i/C)^2 $, and $P[X=0] = 1/C$ 

With a degree of replication $k$ the mean seek distance using the Reimann integral is as follows:

$P[ \mbox{min}( X_1 , X_2 , \ldots , X_k ) \geq i ] = \prod_{j=1}^k P[ X_j \geq i]$	

\vspace{-2mm}
\[
E[X_R] \approx \frac{1}{C^{2k}}\sum_{i=1}^{C-1} (1-\frac{i}{C})^{2k} \approx \int_0^1 (1-x)^{2k} dx = \frac{1}{2k+1},
\]
e.g., for $k=2$ the seek distance is 1/5.

An extension of the analysis in \cite{BiGr88} for writes using the maximum function yields
\vspace{-2mm}
\[
E[X_W] \approx C (1- I_k), \mbox{ where } I_K= \frac{2k}{2k+1} \frac{2k-2}{2k-1} \ldots \frac{2}{3},
\]
For $k=2$ the seek distance is $0.46n$.    

The cumulative distribution and density function 
for the minimum and maximum seek distances for disk systems 
with two independent arms are compared with one-arm systems in \cite{Chie93}. 

This analysis in \cite{BiGr88} does not take into account the effect of R/W heads
being positioned on the same track after a write \cite{LoMa92}, 
so that there is no improvement with shortest seek distance routing, 
i.e., the read seek distance is underestimated and the write seek distance is overestimated. 

The more detailed analysis in \cite{VaMa97b} 
which takes into account disk scheduling and the number of disk cylinders,
shows that previous analyses are accurate for a large number of cylinders and small number of disks.
The same team compares seek distances in mirrored disks without and with cylinder replication
(multiple copies of a cylinder on a disk), 
under various placement policies with uniform and normally distribution of requests \cite{VaMa00} 
It is concluded that cylinder replication yields a significant improvement.

The study of seek optimization in RAID1 configurations 
distinguishes between online and offline algorithms \cite{BaLa05},
Online algorithms know the current position of the disk arm, 
while offline algorithms "choose for each copy $0 \leq i \leq d − 1$ of each file $0 ≤ j ≤ L − 1$ 
a probability $p_{i,j}$ that a read request from file $j$ will be serviced by the $i^{th}$ copy of the file".
Mirrored disks such as ID, GRD, and CD are referred to as semistructured and 
shown to outperform BM as far as seek distances are concerned.
A list of studies for seek optimization is listed on \cite{BaLa05}, 
including previously mentioned $C/5$ based on \cite{BiGr88}, 
which is incorrect in comparison with \cite{CaCF85} discussed below.

Two servers move along $C$ positions on a straight line and in a circle. 
Requests for service from one of the $C$ positions with uniform distribution 
are served from a FCFS queue one at time.
The Nearer-Server (NS) rule is optimal for a circle, 
with an expected normalized server motion $5/36 \approx 0.13889$. 
The average distance over an interval (straight line) is $0.1625$,
while the optimal policy is shown to be is 0.1598, 
so that NS is within 0.1625/0.1598=1.69\% of the optimal policy.
It is better than $C/6 = 0.1666 C$ for the partition rule 
which dedicates servers one server dedicated to $(1,C/2)$ and another to $(C/2+1,C)$.

The {\it Anticipatory Arm Placement (AAP)} in single and mirrored non-zoned disks is presented in \cite{King90}.
With the seek distance in the range (0,1), the arm should be placed at the middle disk cylinder,
reducing the seek distance from 1/3 to 1/4.
In mirrored disks the two arms should be placed at 1/4 and 3/4 
so that the seek distance is 1/8.

AAP in the context of nonzoned and zoned disks with uniform disk accesses 
over all cylinders as well as a provision for hot spots is considered in \cite{ThFu06}.
To simplify the analysis in zoned disks, 
rather than dealing with discrete values for the number of sectors per zone, 
the cylinder/track capacities are set to be proportional to their radius.
%
Denoting the radius of the innermost and outermost cylinder with $R_i$ and $R_o$, 
then $R_i + \sqrt{(R^2_o + R^2_i) / 2} $ divides the disk into two parts with equal capacities,
so that the two arms should be placed at $R_i + \sqrt{( 4 R^2_i + R^2_o ) / 3} $
and $R_i + \sqrt{(2 (R^2_i + 2 R^2_o ) / 3} $.
AAP is detrimental for higher disk utilizations, 
since positioning of the disk arm may delay external requests. 
Simulation results show that there is a crossover point,
so that beyond a certain arrival rate AAP results in degraded performance.

Rearranging data on disk can be used to reduce seek distances,
as discussed for single disks in the Appendix I.
A study that applies cylinder remapping to both single and mirrored disks is \cite{GeRS94}.
Trace analysis of cylinder request streams exhibits strong Markovian dependence
which are used in expressing the mean seek distance for both the single and mirrored disks.
Simulated annealing optimization is used to find permutations reducing seek distance.
The computational costs with a disk with 1760 cylinders is reduced 
by carrying out the optimization on clusters of 40 disk cylinders. 
The optimal permutations for the two disks is non-identical.

\vspace{-5mm}
\section{Mirrored Disk Performance in Various Operating Modes}\label{sec:perf}

We start with a brief introduction to queueing theory applicable to the analysis of RAID performance. 
We analyze RAID performance in normal, degraded, and rebuild modes.

\vspace{-2mm}
\subsection{Queueing Theory for Performance Analysis of Mirrored Disks}\label{sec:queueing}

Disk delay due to disk accesses is a major contributor to mean response transaction time ($R$),
which is the major performance metric for {\it OnLine Transaction Processing (OLTP)} workloads 
with a stringent performance requirement,
e.g., the maximum throughput in {\it Transactions per Second - TPS}
for an OLTP benchmark is determined at $R = 2$ seconds.
Ordinarily $R$ is an increasing function of TPS.

I/O trace analysis of an OLTP workload (an airline reservation system) 
showed that most accesses are to small randomly placed disk blocks \cite{RaBK92},
we hence consider this workload in our discussions.
Access time to large blocks of data is simply determined by disk transfer rate.

Disk accesses can be alleviated by caching, 
e.g., caching the highest levels of a B+-tree alleviates disk accesses. 
Caching methods to improve the performance are discussed briefly in Section~\ref{sec:NVRAM}.
Disk access times can be reduced by appropriate disk scheduling which is discussed briefly in Appendix I.

Our goal is to determine the relative performance of disk array configurations,
rather than estimating accurate performance measures. 
Trace-driven simulation using DiskSim 
\footnote{\url{http://www.pdl.cmu.edu/DiskSim/}.}
used in \cite{WoGP94} are expected to yield accurate results for disks 
whose parameters have been extracted using the DIXtrac tool to extract disk parameters
\footnote{\url{http://www.pdl.cmu.edu/Dixtrac/index.shtml}.}
\footnote{\url{http://www.pdl.cmu.edu/DiskSim/diskspecs.shtml}.}

We adopt the M/G/1 queueing model \cite{Klei75},
because it has been applied in several analytic studies of RAID \cite{ChTo93,ThMe94,MeYu95,ThMe97,ThFH07}.
Disk accesses arrive according to a Poisson process (denoted by M) with rate $\lambda$, 
with exponentially distributed interarrival times with mean $\bar{i} = 1/ \lambda$.
Disk service times are General (G) with $i^{th}$ moment $\overline{x^i}$.
The variance of service time is $\sigma^2_X = \overline{x^2} - (\bar{x})^2$ and 
its coefficient of variation squared $c^2_X = \sigma^2_X / (\bar{x})^2$.
The mean residual service time (following a random arrival) is 
$\overline{x'} = \overline{x^2}/ (2 \bar{x}) $ \cite{Klei75}.
The disk utilization factor is $\rho = \lambda \overline{x} $,
should be less than one to ensure that the queue-length remains finite.
According to Little's result the mean queue-length is $\overline{N}_q = \lambda W $, 
where $W$ is the mean waiting time \cite{Klei75}. 
The mean response time is $R=W + \overline{x}$ With FCFS scheduling the waiting time is independent of service time,
so that the variance of response time is given as follows:
$\sigma^2_R = \sigma^2_W + \sigma^2_X$ \cite{Triv01}.
This equation does not hold for other disk scheduling policies such as SATF.


Given that {\it Poisson Arrivals See Time Averages - PASTA},
the mean waiting time for arrivals with FCFS scheduling is given as follows:
\vspace{-2mm}
\begin{eqnarray}\nonumber  
W = \overline{N}_q \overline{x} + \rho \overline{x'} = \rho W + \frac{\lambda \overline{x^2}}{ 2 \overline{x}} 
\end{eqnarray}
\vspace{-5mm}
\begin{eqnarray}\label{eq:PK}
W = \frac{ \lambda \overline{x^2}}{ 2 (1-\rho) } = 
\frac{\rho \bar{x} (1+ c^2_X)}{ 2 (1- \rho)}.
\end{eqnarray}
For a given $\lambda$ and $\overline{x}$, $W$ increases with $c^2_X$.
For an M/M/1 queueing system with exponential service times: $c^2_X=1$,
$W_E= \rho \overline{x} / (1 - \rho)$ and $R_E = \overline{x} / (1 - \rho)$
and the response time is exponentially distributed: $F_R (t) = 1 - e^{-t/R}$ \cite{Klei75}.
For fixed disk service time $W_F = ( \rho \overline{x}/2)/(1-\rho)$.
For disk service times with $0 \leq c^2_X < 1$ in \cite{ThMe94,ThMe97} $ W_F \leq W \leq W_E$. 
 
Plotting the mean response time $R$ versus $0 \leq \rho < 1 $ by varying $\lambda$.
$R \approx \bar{x} $ for small values of $\lambda$ and 
$R \rightarrow \infty $ as $\lambda \rightarrow 1/ \overline{x}$

For $\rho < 1$ an M/G/1 queue alternates between idle and busy periods 
with means $\overline{g}$ and $\overline{i}=1/lambda$.
Noting that $\rho$ is the fraction of server busy time is: 
$\rho = \overline{g} / (\overline{g} + \overline{i}) $, it follows $\bar{g} = \overline{x}/ (1 -\rho)$.

\vspace{-3mm}
\subsection{RAID1 Performance in Normal Mode}\label{sec:normal}

The arrival rate with to a RAID0/1 disk array with $N=2M$ disks is $\Lambda$.
Assuming that disk loads are balanced due to striping,
the load at each BM pair is $\lambda = \Lambda / M$,
The fraction of read (resp. write) requests is $f_r$ (resp. $f_w=1-f_r$).
The $i^{th}$ moment of read and write accesses is $\overline{x}^i_r$ and $\overline{x}^i_w$.
Read and write accesses to disk have three components: 
seek time, latency and transfer time (see Appendix I). 
Assuming read requests are routed uniformly over the two BM disks
the disk utilization due to read and write requests is:
$\rho_r = \lambda (f_r/2) \overline{x}_r$ and $\rho_w = \lambda f_w \overline{x}_w$.
The maximum arrival rate per disk is $\lambda^{max}_{normal} = [f_r/2 \overline{x}_r + f_w \overline{x}_w]^{-1}$.

The fraction of read and write accesses to disks is 
${f'}_r = (f_r/2) /  (f_r/2 + f_w)$ and ${f'}_w= 1 - {f'}_r$,
so that the $i^{th}$ moment of disk access time is
$\overline{x^i_d} = {f'}_r \overline{x_r^i} + {f'}_w \overline{x_w^i}$
with a mean $\overline{x}_d = {f'}_r \overline{x}_r + {f'}_w \overline{x}_w$ and the mean disk utilization is: 
$\rho = \lambda_d \bar{x}_d = \lambda ( f_r/2 \overline{x}_r + f_w \overline{x}_w ) = \rho_r + \rho_w$, as before.

If read and write requests are processed in FCFS order the mean waiting time is:
\vspace{-2mm}
\[
W = \frac{ \lambda_d \overline{x^2_d} }{ 2 (1-\rho) } .
\]

The response time of read requests can be reduced by processing them at a higher priority than writes.
Given the mean residual service times for reads and writes: 
$\overline{x'}_r = \overline{x^2}_r / (2 \bar{x}_r)$ and $\overline{x'}_w = \overline{x^2}_w / (2 \bar{x}_w)$ 
and disk utilizations $\rho_r$ and $\rho_w$ for reads and writes, 
the mean waiting time for read requests is obtained by applying PASTA again: 
\vspace{-5mm} 
\begin{eqnarray}\label{eq:priority} 
W_r = 
\overline{N^q_r} \overline{x}_r + \rho_r \bar{x'}_r + \rho_w \bar{x'}_w = 
\frac{ \lambda_d \overline{x^2_d} }{ 2 (1 - \rho_r)}, 
\end{eqnarray} 
where we have applied Little's result $\overline{N_q^r} = \lambda W_r$.
The improvement due to prioritizing read requests is: $W_r / W = (1- \rho) / (1- \rho_r)$, 
e.g., 3-fold decrease for $\rho=0.8$ and $\rho_r =0.4 $.


In addition to increasing throughput, mirroring improves the response time for read requests, 
since they can be processed in parallel.
We quantify this effect by assuming that all disk requests are reads ($f_r=1$),
the arrival rate to a mirrored pair is $2 \lambda$,
and disk access time is exponentially distributed with mean $\overline{x}= 1 / \mu$.
With uniform routing with equal probabilities,
each disk is subjected to Poisson arrivals with rate $\lambda$ \cite{Klei75,Triv01}
and the per disk utilization is $\rho = \lambda \overline{x}$.
The mean response time in normal mode is $R_{norm} = \overline{x} / (1- \rho)$
and in degraded mode with one failed disk $R^{degraded} = \overline{x} / (1- 2 \rho) $. 
For $\rho = 0.4$, $R_{degraded} = 5 \overline{x}$ versus $R^{norm} =1.67 \overline{x}$.

{\it Round-robin routing - RR} of read requests results in balanced disk loads with $\rho = \lambda / \mu $ per disk,
but the interarrival times is the Erlang-2 distribution, 
which is the sum of two exponentials with parameter $\lambda$ 
and coefficient of variation $c^2_a =1 /2$ \cite{Klei75}.
Everything else being equal a smaller $c^2_a$ is expected to result in a smaller $W$, 
but this is not generally true and a counterexample is given in \cite{Thom14}.
The mean waiting time for GI/M/1 is: $W_{E-2} = \sigma \overline{x} / (1-\sigma)$ \cite{Klei75},
which is smaller than $W_{M/M/1} = \rho \overline{x} / (1- \rho)$ since $\sigma < \rho$ for $\rho <1$. 
$\sigma$ is the solution to $\sigma = {\cal A}^* (\mu - \mu \sigma)$ 
where ${\cal A}^* (s)$ is the {\it Laplace Stieltjes Transform - LST}.
In the case of the Erlang-2 distribution:
$A^* (s) = [ 2 \lambda / (s + 2 \lambda) ]^2$ \cite{Klei75}.
Setting $s= \mu( 1 - \sigma)$, factoring out $1-\sigma$, we obtain a quadratic equation,
whose negative less than one root is meaningful.
\vspace{-2mm} 
\[
\sigma^2 - (1+ 4 \rho) \sigma + 4 \rho^2=0, \hspace{2mm}\sigma = 
\frac{1}{2} (1 + 4 \rho - \sqrt{1+ 8 \rho})
\]
Applying simple algebra it can be shown easily that $\sigma < \rho$.

Mirrored disks with a shared queue processing read requests can be modeled 
as an M/M/m queueing system, with $m=2$ servers with disk utilizations 
$\rho=2 \lambda \bar{x}/ 2 = \lambda \bar{x}$ and mean response time $R_2 = \overline{x} / (1- \rho^2) $ \cite{Klei75}.
Setting $\rho=0.9$ $R_2 \approx 5 \overline{x}$, 
while $R_1= 10\overline{x}$ for a single disk.

To attain lower response times, it is meaningful at times 
to issue a read request to both disks and accept the first completed request.
Disks modeled as M/M/1 queues have exponential response times as noted earlier, 
so that $R_{min} = R/2$ \cite{Triv01}.

The completion time of writes on both mirrored disks represented as M/M/1 queues is:
$R_2^{F/J} = R_1 (1.5 - \rho/8)$ \cite{Thom14},
which is smaller than the maximum of two response times $R_2^{max} = H_2 R$,
where $H_k = \sum_{i=1}^k 1 / i $ is the Harmonic sum \cite{Triv01}.
This mean response time holds if all requests are writes,
but simulation results have shown that with interfering read requests 
the write response time is higher and closer to $R_2^{max}$, when the overall $\rho$ remains the same.

Performance analyses of RAID1 in normal and degraded mode are given in \cite{ChTo96,ThXu08}.
The performance of LSI RAID with RMW versus 
{\it Reconstruct Write (RCW)} method \cite{Thom05a} is given in \cite{ThTa12}.

\vspace{-2mm}
\subsection{Operation in Degraded Mode}\label{sec:degraded}

When one of two mirrored disks fails in BM the read load on the surviving disk is doubled.
The mean disk service time is $\overline{x}_d^{deg} = f_r \bar{x}_r + f_w \bar{x}_w$
and $\lambda_{max}^{deg} = 1 / \overline{x}_d^{deg}$.
The maximum requests per second drops 2-fold for $f_r=1$, 
but this drop is less significant for a large $f_w$.
RAID1/0 tolerates $k \leq M$ disks failures and assuming balanced disk loads due to striping: 
$R_r^{overall} = (k/M) R_r^{deg} + (1-k/M) R_r^{norm}$.

In the case of ID with $n$ disks we assume the read load 
is evenly distributed between primary and secondary data blocks,
so so that the normalized load at each disk is normal mode is one, since:
$\frac{1}{2} + (n-1) \times \frac{1}{2(n-1)} = 1$.
When a disk fails the load on remaining disks increases by $1/(n-1)$, so that: 
$1 + \frac{1}{n-1} = \frac{n}{n-1}$, 
so that for ID $n=4$ the load increase is 33.3\%, versus 50\% for BM.

Denoting the load in normal mode on primary and secondary disks in GRD by one, 
the load of secondary disks increases by $k/M$ when $k$ primary disks fail.
We reduce the load on primary disks with $k$ disk failures a fraction $1- \alpha_k$ of read requests,
so that a fraction $\alpha_k$ is routed to secondary disks.
To balance disk loads on both sides we set $1-\alpha_k = \alpha_k +k/M$, hence $\alpha_k = (M-k)/2M$,
so that with $k=M$ failures the load on primary disks is zero and secondary disks receive twice the load. 

Static routing of requests can be used to balance disk loads 
due to reads in CD as illustrated in Figure~\ref{fig:CD'} \cite{HsDe90} in Section{sec:org}.
Since CD can tolerate multiple disk failures the routing probabilities should be adjusted,
as shown by the last two rows of the figure.

As a simple example we consider balanced disk loads in normal mode with $f_r=1$.
$D_3$ load will be doubled while the loads on remaining disks will increase to $6/5$.
In case $D_7$ fails the loads of $D_5$-$D_6$ and $D_8$-$D_1$ will increase to 3/2 of normal load.
This is an instance when given a single spare disk the rebuilding 
$D_2$ or $D_4$ should be prioritized with respect to $D_7$.



\vspace{-2mm}
\subsection{Rebuild Processing in RAID1}\label{sec:rebuild}

Both RAID1 and RAID5 are one disk failure tolerant (1DFTs) 
and are susceptible to data loss if a second disk fails,
so a spare disk should be provided to start rebuild immediately if a spare disk is available,
Disk failures prediction by predictive failure analysis reduces rebuild overhead 
by copying the contents of a failing disk onto a spare disk. 
\footnote{\url{https://www.computerworld.com/article/2846009/the-5-smart-stats-that-actually-predict-hard-drive-failure.html}.}
Rebuild time can be minimized if the processing of external disk requests is quiesced, 
but this is unacceptable because the cost of downtime is quite high 
for most lines of business as shown in Figure 1.3 in \cite{HePa07}, 
so rebuild processing is carried out while the disk array operates in degraded mode.

Rebuild processing in RAID5 XORs consecutive {\it Rebuild Units - RUs},
which may be tracks to reconstruct the contents of the failed disk on a spare disk.
Repair with rebuild processing affects the mean response of disk accesses 
as quantified in Section~{sec:rebuildana}.
A subset of disks in HDAs need to be XORed to reconstruct data,
but this step is not required for RAID1, 
which simply copies the contents of the surviving disk onto a spare.  

The time to read the contents of an idle disk: $T_{copy} (0) = N_{track} \times T_{rot}$,
where $N_{track}$ is the number of disk tracks and $T_{rot}$ is the disk rotation time.
We ignore minor delays such as track and cylinder skew \cite{JaNW08}.
Given that the disk utilization due to user requests in normal mode was $\rho$, 
the disk utilization in degraded mode with all read requests is $\rho^{deg}= 2 \rho$.
and the fraction of time that the surviving disk is idle and available rebuild processing is $1-2 \rho$.

The {\it Vacationing Server Model (VSM)} \cite{Taka91} 
utilized in \cite{ThMe94} extends the analysis in \cite{Meno94} 
to analyze RAID5 performance in normal, degraded, and rebuild modes 
with an M/G/1 rather than M/M/1 queueing model with general 
rather than exponential disk service times utilized in \cite{Meno94}.
The analysis quantifies the effect of rebuild processing 
on the mean response time of external disk requests and also determines the rebuild time,
which can be approximated by the time to read a surviving disk, since disk loads are balanced.


Rebuild requests are processed at a lower priority than external requests when the queue of external requests is emptied,
at which time the server starts taking vacations, 
which correspond to reading successive RUs/tracks from the surviving disk.
There are two types of vacation: 
type one vacations ($V_1$) require a seek to access the next track to be rebuild and a disk rotation to read the track.
Successive type 2 vacations ($V_2$) read consecutive tracks without incurring seek.

While the analysis in \cite{ThMe94} takes into account two vacation types,
we simplify the discussion by ignoring the seeks required for $V_1$ vacations 
and consider only $V_2$ rebuild requests,
whose residual vacation time is $\overline{v}_r = \overline{v^2} / ( 2 \bar{v}) =T_{rot}/2$, 
i.e., half of disk rotation time.

We ignore the difference between the access time of read and write requests to randomly placed disk blocks: 
$\overline{x}_r \approx \overline{x}_w = \overline{x}$.
Given that the processing of first request is delayed by $\overline{v}_r$,
then the first disk request has duration $\overline{y}= \overline{x} + \overline{v}_r$ 
and the modified busy period, which is called a delay cycle $\overline{d} = \overline{y} / (1-\rho)$.

We use an argument similar to the one used in obtaining $\overline{g}$ to obtain the delay busy period $d$.
\vspace{-3mm}
\begin{eqnarray}\label{eq:dcp}
\rho =
\frac{\overline{d} - \overline{v}_r}
{\overline{d} + 1/\lambda}
\hspace{2mm}
\Longrightarrow
\hspace{2mm}
\overline{d} =
\frac{\overline{x} + \overline{v}_r } { 1- \rho}.
\end{eqnarray}

The mean time to the arrival of the first request 
after the completion of a busy period is $1/\lambda$ 
so that we define a cycle as $T_{cycle} = \overline{g'} + 1/\lambda$.
The probability that a request arrives during rebuild is $p=\lambda T_{rot}$ \cite{Klei75},
so that the distribution of the number rebuilds per idle period is 
$P_n= (1- p) p^{n-1}$ with a mean $\overline{n}=1/p$.
Given a disk with $N_{track}$ tracks, 
rebuild time is $T_{rebuild} = N_{track} \times T_{cycle} / \overline{n}$.

With read redirection in effect reconstructed data blocks are read directly 
from the spare disk rather than reconstructing them on demand \cite{MuLu90}.
This results in a reduction in disk loads in degraded mode processing and an acceleration of rebuild processing.
In the case of RAID1 the read load on the surviving disk is reduced 
as read requests to reconstructed blocks are directed to the spare disk.
As rebuild progresses in addition to redirected reads 
data blocks reconstructed on the spare disk should be updated.
The fraction of read requests to be redirected to accelerate rebuild rate is determined in \cite{MuLu90}.
Rebuild analysis should be carried out in stages 
to take into account the variation in rebuild load \cite{ThMe94,ThMe97} 

To estimate the mean waiting while is in progress we apply the PASTA principle again: 
$W_{VSM} = \bar{N}_q \bar{x} + \rho \bar{x'} + (1-\rho) \bar{v'}$ and simplifying:
\vspace{-2mm}
\begin{eqnarray}\label{eq:VSM}
W_{VSM} = 
\frac{ \lambda \overline{x^2}} { 2 (1- \rho)} + \frac{\overline{v^2}}{2 \overline{v}} 
= W_{M/G/1} + \frac{ \overline{v^2} }{ 2 \bar{v}}.
\end{eqnarray}
This corresponds to Eq. (2.14a) in Chapter 2 in \cite{Taka91}.
Eq. (11) for $W_{VSM}$ in \cite{BaSc02} uses Eq. (2.40a) in Chapter 2 in \cite{Taka91},
which is for the case when the first request has an exceptional service time.
This is not the case here.

Queues with the {\it Permanent Customer Model (PCM)} process two types of requests:
(i) ordinary requests to randomly placed disk blocks with Poisson arrivals which are served in FCFS order,
(ii) permanent requests which rejoin the tail of the queue 
of ordinary requests upon completing service and are also served in FCFS order \cite{BoCo91}.
PCM was adopted to analyze rebuild in mirrored disks in \cite{MeYu94}.
where permanent requests represent the reading of a track, 
while ordinary requests are accesses to read and update randomly placed disk blocks.
PCM and VSM rebuild in ordered and greedy order in mirrored disks are compared in \cite{BaSc02}
Greedy rebuild is intended to reduce rebuild time by out-of-order processing of track reads,
i.e., by reading the closest unread track to the read-write head after the last ordinary request is processed.
Note that greedy rebuild is not applicable to RAID5, 
because of excessive buffer space requirements.

VSM incurs fewer seeks than PCM in reading successive tracks,
since the probability that there are interim arrivals and 
the R/W head moves after reading the current track is lower for VSM than PCM:
$P_{VSM}= 1 - e^{- \lambda T_{rot}}$ and $P_{PCM} = 1 - e^{\lambda ( T_{rot} + N_q \overline{x} )}$,
where $T_{rot}$ is track rotation time.
As shown in Figures 3(a) and 3(b) rebuild time with VSM is lower than PCM, 
and except for the highest arrival rates $R_{PCM} > R_{VSM}$.
The greedy policy shows very little improvement for $R_{VSM}$, 
which significantly outperforms $R_{PCM}$,
this is because VSM processes rebuild requests at a lower priority than external requests,
while PCM processes them at the same priority.



Rebuild processing can be parallelized with the CD, ID, and GRD RAID1 organizations.
The rebuilding of a failed disk $D_i$ for the CD organization 
can be parallelized by copying from $D_{i-1}$ and $D_{i+1}$ into $D_i$.
$n-1$-fold parallelism is possible with the ID organization for rebuilding primary data.
In the case of GRD with $N=2M$ disks, $M$ disks can participate in rebuild processing.
At low disk utilization the spare disk will constitute a bottleneck
and the reading rate from source disks should be throttled.

\vspace{-5mm}
\section{Reliability Analysis}\label{sec:rel}

We first provide reliability expressions for RAID1 organizations 
with no repair and obtain their MTTDL \ref{sec:conv}. 
In Section~\ref{sec:approxRel} we discuss approximate reliability analysis,
which is suited for the comparing of the relative reliabilities 
of RAID1 configurations against each other and RAID$(4+k)$ arrays.
Data loss may occur due to disk controller failures and an analysis using 
{\it Continuous Time Markov Chains - CTMCs} is discussed in Section~\ref{sec:controller}.
The MTTDL of RAID1 with repair is obtained in Section~\ref{sec:rebuildana}.
Storage reliability modeling research at IBM Research at Zurich,
which is relevant to this study is reviewed in Section \ref{sec:Zurich}.
In response to articles that MTTDL is no longer a good measure of reliability,
especially for RAID6, we refer the reader to \cite{IlVe15b},
which gives a solid response this criticism and continue using the MTTDL. 
We conclude with simulation techniques reliability modeling in Section \ref{sec:sim}.

Disk failure rates are reported in \cite{Gibs92} and \cite{ScGi07}, 
which is a followup study that takes into account other storage components.
The frequency of {\it Latent Sector Errors - LSEs} is reported in \cite{ScDG10}.
which also investigates the effectiveness of disk scrubbing and 
{\it IntraDisk Redundancy (IDR)} in dealing with LSEs.
The weaknesses of \cite{ScDG10} are addressed in \cite{IHHE11}.

\vspace{-2mm}
\subsection{Reliability Expressions for Mirrored and Hybrid Disk Arrays}\label{sec:conv}

The reliability function of a system is the complement of 
the probability distribution of time to failure: 
$R(t)=1-F(t)$ \cite{Triv01}.
In our discussion we will be mainly concerned with disk reliabilities
rather than DACs, interconnects, power supplies, cooling fans, etc,
since we are interested in the relative reliability of various RAID configurations,
rather than assessing the overall reliability. 

The Weibull distribution \cite{Triv01} is a good fit 
for the time to disk failure according to \cite{Gibs92,ScGi07},
but most mathematical analyses of RAID reliability use the exponential distribution 
$R(t) = e^{-\delta t}$ as an approximation to the Weiball distribution 
because of its mathematical tractability \cite{Gibs92,Triv01}.
The {\it Mean Time to Failure (MTTF)} of a disk is 
$\mbox{MTTF}_{disk} = \int_0^\infty R(t) dt = 1 / \delta$.
One consequence of the exponential distribution is that the reliability of $n$ disks is:
$R_{n-disks} (t) = R^n (t) = e^{- n \delta}$ and $\mbox{MTTF}_{n-disk}  = \mbox{MTTF}_{disk}/n $.
This is a justification for introducing redundancy in large disk arrays.

CTMC models are applicable to modeling the failure and repair process in disk arrays \cite{Triv01}.
While repair time is not exponential, 
analytical models of repair have assumed an exponential distribution for tractability.
The Proteous simulator described in \cite{KPLS13} showed little difference between values obtained 
assuming deterministic versus exponential repair time, which is required by CTMC models.
This simulator was used to predict the risk of data loss 
for RESAR disk array \cite{Sch+16} described in Section~\ref{sec:org}.
We start the discussion with the no repair case, which can be modeled as a pure death-process.

Let $\boldsymbol{{\cal S}}_i$ denotes the state with $0 \leq i \leq I$ failed disks,
where $I$ is the maximum number of disk failures that can be tolerated without data loss.
Let $A(N,i)$ denote the number of possibilities that $i$ disk failures do not lead to data loss.
The probability that $i^{th}$ disk failures does not lead to data loss is then:

\vspace{-2mm}
\[
p_i = A(N,i) / {N \choose i}, 1 \leq i \leq I.
\]
The rates of transitions not leading to data loss for failed disk are: 
$\boldsymbol{{\cal S}}_{i-1}$ to $\boldsymbol{{\cal S}}_{i}$: $(N-i+1) p_i \delta$.
The rates of transitions leading to the failed state ($\boldsymbol{{\cal S}}_F)$ are 
$(N-i+1)q_i \delta$, $2 \leq i \leq N$, where $q_i = 1 - p_i$.

For RAID1 with $N=2M$ disks, $I=M$ for BM and most RAID1 organizations, 
and $I=c$ for the ID organization, since only one disk failure per cluster is allowed.
Generally, $A(N,i)= {N \choose i}, i=0,1 $ and $A(N,i)=0$ for $i > I$.
The expressions for RAID reliability with the setting $r=R(t)$ can be expressed as:

\vspace{-7mm}
\begin{eqnarray}\label{eq:rel}
R_{RAID} (N) = \sum_{i=0}^I A(N,i) r^{N-i} (1-r)^i .
\end{eqnarray}

In the case of RAID1/0 with $N=2M$ disks (RAID1 at lower level and RAID0 at higher level),
which is a RAID0 array with $M$ virtual disks each one of which is a BM,
as few as two disk failure may lead to data loss,
but up to $M$ disk failures can be tolerated, 
as long as failed disks are not pairs.

\vspace{-5mm}
\begin{eqnarray}\label{eq:BM}
A (N,i) = { M \choose i } 2^i , \hspace{2mm} 0 \leq i \leq M.
\end{eqnarray}
For $n$-way replication with $M= N/n$ groups of $n$-way replicated disks:
$A (N,i) = { M \choose i } n^i , \hspace{2mm} 0 \leq i \leq M$.

In the case of GRD and RAID0/1 up to $M$ disks on either side 
can fail as long as they are all one side: 

\vspace{-5mm}
\begin{eqnarray}\label{eq:GRD}
A (N,i) = 2 { M \choose i } , \hspace{2mm} 0 \leq i \leq M.
\end{eqnarray}
ID with $c$ clusters and $n=N/c$ disks per cluster, 
can have only one disk failure per cluster and any one of the $n$ disks in a cluster can fail.

\vspace{-5mm}
\begin{eqnarray}\label{eq:ID}
A (N,i) = { c \choose i } n^i , \hspace{2mm} 0 \leq i \leq c.
\end{eqnarray}
The expression for $A(N,i)$ for CD is derived in \cite{ThBl06} and also in the Appendix in \cite{GaBi00}:

\vspace{-5mm}
\begin{eqnarray}\label{eq:CD}
A (N,i) = { {N-i-1} \choose {i-1} } + {{N-i} \choose i}, \hspace{2mm} 1 \leq i \leq M. 
\end{eqnarray}

Let $v_i$ denote the number of visits to state $\boldsymbol{{\cal S}_i$.
When repair returning the system to its initial state is not allowed $v_0=1$ for $\boldsymbol{\bf S}}_0$.

\vspace{-7mm}
\begin{eqnarray}\label{eq:visits}
v_i = v_{i-1} p_i,  \hspace{2mm} v_i = \prod_{j=0}^i p_i, \hspace{2mm} 1 \leq i \leq I.
\end{eqnarray}

A closed form expression for $A(N,i)$ for LSI Logic RAID is not available, 
but can be obtained using enumeration \cite{ThTa12}.
$A(8,i) = {8 \choose i}, 0 \leq i \leq 2$, 
$A(8,3)={8 \choose 3} - 4= 52$ 
since there are four cases where the Ddisk is in the middle, so that $q_3=4/56=1/14$.

There are ${8 \choose 4}$ configurations of four disk failures,
but according to the enumeration given in Table 2 only 25 cases lead to data loss, so that $q_4 = 25/70 = 5/14$,
which differs from $q_4 = 16/65$ given in Section 3.1.2 in \cite{ALPS08}.

The reliability analyses of SSPiRAL(4+4,3) disk array in \cite{ALPS08} uses a CTMC,
which yields the reliability expression \cite{Triv01},
which can be used to determine the probability of data loss at the end of its "economic lifespan". 
We have the following transitions ${\cal S}_i \rightarrow {\cal S}_{i+1}, 
\mbox{ with failure rate } (8-i) \lambda,  0 \le1 i \leq 3$ and
${\cal S}_{i+1} \rightarrow {\cal S}_i, 0 \le1 i \leq 3 \mbox{ with repair rate } i \mu, 2 \geq i \geq 0$.
Finally, $(1/5)^{th}$ of failures with rate $5\lambda \mbox{ from } {\cal S}_3$ lead 
to data loss (${\cal S}_{DL}$) and $(4/5)^{th})$ to ${\cal S}_4$. 
There are two transitions from ${\cal S}_4$: 
with rate $4 \lambda \mbox{ to }{\cal S}_{DL}$ and $4 \mu\mbox{ to }{\cal S})_3$.

The analysis has the following shortcomings:
(i) The repair rate is set proportional to the number of failed disks
and does not take into account potential hardware bottlenecks.
(ii) An infinite supply of spare disks is postulated.
(iii) Repair rates are exponentially distributed,
and as a new disk fails due to the memoryless property of this distribution
it is as if the rebuild process at all disks under repair is restarted.
(iv) The analysis does not take into account LSEs,
which are the main cause of rebuild failures \cite{IHHE11}.

SSPiRAL(4+4,3) tolerates up to three failed disks. 
Eighty percent  of four disk failures are tolerated for $N=8$,
so that $v_4 = 4/5$ and $A(8,4) = (4/5) {8 \choose 4} =56$.

The reliability of RAID$(4+k)$ arrays with $N$ disks,
which can tolerate up to $k$ disk failures is: 

\[
R_{RAID(4+k)} (N) = \sum_{i=0}^k  {N \choose i} r^{N-i}  (1 - r)^i.
\]

\begin{figure}[htb] 
\centering
\includegraphics[scale=0.650]{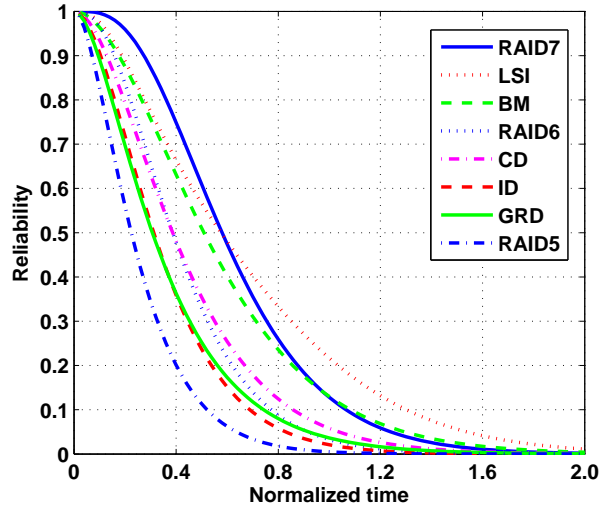}\\                   
\caption{\label{fig:reliability}Reliability versus time normalized with respect to disk MTTF \cite{ThTa12}.}
\end{figure}

In Figure~\ref{fig:reliability} we plot the reliabilities
for various RAID1 organizations and RAID5/6/7 and LSI RAID for $N=8$ disks 
versus time normalized with respect to MTTF.
For small values of $t$, 
RAID7 which tolerates all three disk failures has the highest reliability,
while LSI is second since it tolerates half of three disk failures.
For smaller values of $t$ RAID6 is more reliable than BM,
which is the most reliable of mirrored disk organizations,
but there is a crossover point since BM can tolerate more than two disk failures.
RAID5 has the lowest reliability since it can only tolerate single disk failures.  

The MTTDL in a RAID systems can be obtained by integration
or noting that the mean holding time: $H_i = [(N-i)\delta]^{-1}$ 
is the inverse of the failure rate at $\boldsymbol{{\cal S}}_i$:

\vspace{-5mm}
\begin{eqnarray}\label{eq:MTTDL}
MTTDL= \int_0^{\infty} R_{RAID} (N) dt=
\sum_{i=0}^M v_i H_i =
\sum_{i=0}^M \frac{v_i}{(N-i)\delta}.
\end{eqnarray}

A better measure than the MTTDL of a RAID system is its reliability $R_{RAID} (t_u)$,
where $t_u$ is its useful lifetime.
To justify this point consider a single system 
with failure rate $r(t)=e^{delta}$ with $\mbox{MTTF}_{single} = 1/\delta$,
while in the case of {\it Triple Modular Redundancy - TMR} system,
where at least 2-out-of-3 components are required for the correct operation of the system:
$R_{TMR} (t) = r^3 (t) +3r^2 (t) (1-r(t))$ and $\mbox{MTTF}_{TMR} = 5/ (6 \delta)$. 
Setting $R_{TMR} (t) = R_{single} (t)$ yields the crossover point $t_u=ln {2}) / \delta \approx 0.693 / \delta$,
and $R_{TMR} (t) > R_{single} (t) \mbox{ for }t < t_u$.

\begin{table*}
\renewcommand\arraystretch{1.5}
\centering
 \begin{tabular}{|c|c|c|c|c|c|c|c|c|c|}
 \hline
 RAID5 & BM & CD & GRD & ID & RAID6 & LSI & RAID7 & SSP & R8\\
 \hline
 $\frac{15}{56 \delta}$                          
& $\frac{163}{280 \delta}$                       
& $\frac{379}{840 \delta}$                       
& $\frac{3}{8 \delta }$                          
& $\frac{61}{168 \delta}$                        
& $\frac{ 73}{ 168 \delta}$                      
& $\frac{521}{840 \delta}$                       
& $\frac{638}{840 \delta}$                       
& $\frac{701}{840 \delta}$                       
& $\frac{743}{840 \delta}$   \\                  
 \hline
  \scriptsize {$0.268 \delta^{-1}$}              
& \scriptsize {$0.582 \delta^{-1}$}              
& \scriptsize {$0.451 \delta^{-1}$}              
&  \scriptsize {$0.375 \delta^{-1}$}             
& \scriptsize {$0.363 \delta^{-1}$}              
& \scriptsize {$0.435 \delta^{-1}$}              
& \scriptsize {$0.620 \delta^{-1}$}              
& \scriptsize {$7595 \delta^{-1}$}               
& \scriptsize {$0.8345 \delta^{-1}$}             
& \scriptsize {$0.884   \delta^{-1}$}  \\        
 \hline
${N \choose 2} \varepsilon^2$                    
& $\frac{N \varepsilon^2}{2}$                    
& \scriptsize{$N\varepsilon^2$}                  
&  $\frac{N(N-1) \varepsilon^2}{4}$              
& $\frac{N(N-c) \varepsilon^2}{2c}$              
& $ {N \choose 3} \varepsilon^3$                 
& $ \frac{N}{2}\varepsilon^3$                    
& ${N \choose 4} \varepsilon^4$                  
& $ \frac{1}{5}  {8 \choose 4} \varepsilon^5$    
& $ {N \choose 5} \varepsilon^5 $  \\            
\hline
\end{tabular}
\caption{\label{tab:MTTDL}MTTDLs as a ratio and a fraction of the MTTF ($\delta^{-1}$)
and the first term in asymptotic reliability expression with $\epsilon$ denoting the unreliability.}
\end{table*}

MTTDLs for RAID1, RAID5-8, LSI, SSPiRAL and Weaver codes are given in Table \ref{tab:MTTDL}.
RAID5 arrays are less reliable than mirrored disks since they cannot tolerate more than one disk failure.
Mirrored disks in increasing order of their MTTDL are ID, GRD, CD, BM.
LSI Logic RAID is more reliable than RAID6 because it tolerates all two disk failures, 
but also one half of three disk failures.
SSPiRAL(4+4,3) tolerates all three disk failures and most four disk failures, including all Ddisk failures.
Weaver(8,3,3) is less reliable than RAID8, since it does not tolerate all four disk failures.

\vspace{-2mm}
\subsection{Performability Analysis}

The performability measure combines the failure process with performance \cite{Triv01}, 
In \cite{ThXu08} we define performability as the number of 
I/O requests that are processed by a disk array system 
from the beginning of its operation to the point that data loss occurs.
Using a birth death process the performability is obtained 
by summing over all intervals with varying number of failed disks ($i$)
the product of the durations of the interval times 
and the maximum throughput denoted by $T_{N-i}, 0 leq i \leq I$, 
e.g., $T_N = N/ \overline{x}_d$, 
where overline $\overline{x}_d$ is the disk service time:
$$
{\cal P} = \sum_{i=0}^I \frac{v_i T_i}{ (N-i) \delta}.
$$
Note similarity to Eq.~(\ref{eq:MTTDL}) to compute the MTTDL.
The performability of various RAID organizations is compared via barcharts in \cite{ThXu08}.

\vspace{-2mm}
\subsection{Approximate Reliability Analysis}\label{sec:approxRel}

The reliability analysis in \cite{Thom06b} expresses disk reliability as $r=1 - \epsilon $, 
where $\epsilon \ll 1$.
For example, assuming an exponential distribution and an MTTF$=10^6$ hours or 114 years,
after three 3 years $R(3) \approx 1 - 3/114 = 0.975$ and $\epsilon = 0.025$.
The reliability of a system is then expressed as one minus the smallest power of $\epsilon$.
Note that the exponential distribution is not required for this analysis.
For example, for $n$-way replication has data loss with $n$ disk failures and allows at most $n-1$ disk failures:
$ R_{n-way} (n) = 1- (1-r)^n \approx 1- \epsilon^n$. 

For the BM configuration we utilize the expression for $A(N,i)$ given by Eq.~(\ref{eq:BM}),
but only retain the $\epsilon^2$ term, which is sufficient for comparing reliabilities.
\footnote{We have corrected the first part of Eq.~(7) in \cite{Thom06b}.}

\vspace{-2mm}
\[
R_{BM} (N) = \sum_{i=0}^{N/2} { N/2 \choose i} 2^i r^{N-i} (1-r)^i \approx 
r^N + Nr^{N-1} r^{N-1} (1-r) + 4 {{N/2} \choose 2} r^{N-2} (1-r)^2 \approx 1- \frac{N}{2} \epsilon^2 .
\]

For GRD or RAID0/1 instead of using $A(N,i)$ in Eq.~(\ref{eq:GRD}),
we consider an alternative method.
Each of two mirrored virtual disks consists of $M$ disks and $R_M = r^M$, so that:

\vspace{-5mm}
\begin{eqnarray}\label{eq:apprGRD}
R_{GRD} (N) =  1- (1- R_M)^2 = 2 r^M - r^{2M} \approx  1 - \frac{N(N-1)}{4} \epsilon^2 .
\end{eqnarray}
In the case of ID we use the expression for $A(N,i)$ given by Eq.~(\ref{eq:ID}).

\vspace{-5mm}
\begin{eqnarray}\label{eq:apprID}
R_{ID} (N) = r^N + Nr^{N-1} (1-r) + \frac{c(c -1)}{2}n^2 r^{N-2} (1-r)^2
\approx 1 - \frac{(N(N-1) - c(c-1)n^2}{2} \epsilon^2 .
\end{eqnarray}
For $c=N/2$ ID is equivalent to BM.

In the case of CD we do not need the expression for $A(N,i)$,
but argue that out of ${N \choose 2}$ two disk failures,
there are $N$ consecutive two disk failures that lead to data loss, 
so that $A(N,2) = N(N-1)/ 2 - N = N(N-3)/2$.

\vspace{-5mm}
\begin{eqnarray}\label{eq:apprCD}
R_{CD} (N) \approx r^N + N r^{N-1} (1-r)+ \frac{N(N-3)}{2} r^{N-2} (1-r)^2 
\approx 1 - N \epsilon^2.
\end{eqnarray}

Note that BM is the most reliable among the four mirrored disk organizations followed by CD,
while ID is more reliable than GRD since $N(2/c -1) < 1$.

LSI tolerates all single and double disk failures:
$A(N,i)= {N \choose i}, 0 \leq i \leq 2$,
but only half of three adjacent disk failures, 
so that $A(N,3) = {N \choose 3} - N/2 $. We have:

\vspace{-5mm}
\begin{eqnarray}\label{eq:LSI}
R_{LSI} (N) \approx 1 - \frac{N}{2} \epsilon^3.
\end{eqnarray}

\vspace{-5mm}
\begin{eqnarray}\label{eq:SSP}
R_{SSP} (8) \approx 1 - 14 \epsilon^5.
\end{eqnarray}

Disk arrays with erasure coding are described in Appendix II.
In the case of the $k$DFT RAID$(k+4)$, $k \geq 1$, 
the asymptotic reliability expression is given in \cite{ThTH12}:

\begin{equation}\label{eq:appr}                                     
R_{RAID(4+k)} \approx 1 - {N \choose {k+1}} \epsilon^{k+1} + (k+1) { N \choose {k+2}} \epsilon^{k+2} - \ldots,
\end{equation}
but only the first term, which is negative, is required in approximations
In what follows we set $N=8$, which yield 
$R_{RAID5} (8) = 1 - 28 \epsilon^2$,
$R_{RAID6} (8) = 1 - 56 \epsilon^3$,
$R_{RAID7} (8) = 1 - 70 \epsilon^4$,
$R_{RAID8} (8) = 1 - 56 \epsilon^5$,

The following conclusions can be drawn for $N=8$.

\begin{enumerate}

\item
RAID5 is less reliable than the four RAID1 organizations,
since RAID1 arrays can tolerate more than two disk failures up to $M=N/2$.

\item
RAID6 is less reliable than LSI RAID, since the latter in addition 
to all two disk failures tolerates half of three disk failures,
but RAID7 is more reliable than LSI RAID since it tolerates all three disk failures,
while LSI RAID does not..
\vspace{-3mm}
$$
R_{LSI} (N) \approx 1 - \frac{N}{2} \epsilon^4,
$$

\item
RAID8 is more reliable than SSPiRAL(4+4,3), since the latter cannot tolerate $1/5^{th}$ of four disk failures. 
\vspace{-3mm}
$$
R_{SSPiRAL} (8) \approx 1 - 14  \epsilon^5.   
$$

\end{enumerate}
The above discussion is summarized in Table~\ref{tab:MTTDL}.
\vspace{-2mm}
\subsection{RAID Reliability with Repair}\label{sec:rebuildana}

Early analyses of RAID reliability with repair dealt with whole disk failures \cite{Gibs92}.
These analyses were later extended to include the effect of LSEs,
but also intradisk redundancy and disk scrubbing on LSEs \cite{IHHE11}.
Adjusting the disk failure rate has been used to rake account these complications.

The expression for reliability $R(t)$ 
can be obtained by solving the related linear differential expressions.
The analysis in \cite{Gibs92} yields the reliability $R(t)$ of RAID5 
as the sum of two exponentials \cite{Gibs92}.
The MTTDL can be obtained by integrating $R(t) = 1 - F(t)$: $\mbox{MTTDL}=\int_0^\infty R(t) dt$.
Specialized packages {\it SHARPE—Symbolic Hierarchical Automated Reliability and Performance Evaluator
- SHARPE} \cite{Triv01} were applied in \cite{Gibs92} to analyze more complex cases.

A shortcut method to obtain the RAID5 MTTDL with $N$ disks,
failure rate $\delta$, and repair rate $\mu$ which provides insight into the rebuild process.
Operation of the systems is specified by a 3-state CTMC:
${\cal S}_{N}$: no failed disks,
${\cal S}_{N-1}$: single failed disk,
${\cal S}_{N-2}$; two failed disks and data loss.
The transition rates are
${\cal S}_{N} \rightarrow {\cal S}_{N-1}$: $N \delta$,
${\cal S}_{N-1} \rightarrow {\cal S}_{N-2}$: $(N-1) \delta$,
${\cal S}_{N-1} \rightarrow {\cal S}_{N}$: $\mu$.
At ${\cal S}_{N-1}$ there are two competing transitions with exponential rates:
rebuild is successful the repair process finishes before another disk fails,
otherwise there is data loss.
The probability of a successful rebuild is the ratio of transition rates
$p = \mu /(\mu + (N-1) \delta )$ \cite{Triv01},
The distribution of successful rebuilds is
$P_{succ} = (1-p_s) p_s^k , \hspace{1mm} k \geq 0 $, which has a mean
$v_N =  {p}/ ({1-p}) = \mu/ N \delta = MTTF / ( (N-1) \times MTTR )$.
The MTTDL is dominated by the time spent at ${\cal S}_N$,
which is exponentially distributed with mean holding time $H_N =MTTF/N$,
while the time spent in ${\cal S}_{N-1}$ is small, since $ \mu \gg (N-1) \lambda$/
Multiplication by $v_N$ yields:
\vspace{-3mm}
\begin{eqnarray}\label{eq:MTTDLR5}
\mbox{MTTDL}_\infty \approx v_N H_N = \frac{\mu}{N (N-1) \lambda^2} = \frac{ \mbox{MTTF}^2 }{ N (N-1) \mbox{MTTR}}.
\end{eqnarray}
The subscript specifies an unlimited number of spare disks.
For $N=2$ we have the MTTDL for RAID1.

The analysis in \cite{Dho+08} yields 

\[
\mbox{MTTDL}= \frac{ (2N-1) \lambda+ \mu}{N \lambda [ (N-1) \lambda + \mu P_{uf}}
\]
which extends the analysis of \cite{Gibs92} 
by multiplying $\mu$ with $P_{uf}$, which is the probability of an uncorrectable disk failure. 

The analysis in \cite{Che+94} provides the MTTDL of RAID6 arrays as 
\[ 
\mbox{MTTDL}_2 =\frac{ \mbox{MTTF}^3 }{ N(N-1)(N-2) \mbox{MTTR}^2 } ,
\]
which assumes that the repair-rate with two disk failures remains $\mu$.
If RAID$(4+k)$ tolerates $k$ failures a generalization of the above formulas is:
\[
\mbox{MTTDL}_k = \frac{ \mbox{MTTF}^{k+1} (N - k -1) }{ N! \mbox{MTTR}^k }
\] 

Unlimited repairman were postulated in \cite{Angu88} and his analysis yields

\vspace{-3mm}
\[
\mbox{MTTDL} \approx 
\frac{ \mbox{MTTF}^{k+1} } { (N-k) {N \choose k} \mbox{MTTR}^k }
\times \sum_{i=0}^k {N \choose i} \left( \frac{ \mbox{MTTR}} {\mbox{MTTF}} \right) ^i
\approx \frac{ \mbox{MTTF}} { (N-k) \times {N \choose k} } \times 
\left( \frac{\mbox{MTTF}} {\mbox{MTTR} } \right) ^k.
\]
It is easy to see that the MTTDL with the latter formula estimates an MTTDL which is $\approx k!$
times higher than the original MTTDL formula given in \cite{Che+94}. 

A more realistic MTTDL can be obtained by allowing a few spares.
In a RAID5 disk array with $N$ disks, one of which is a spare,
we have four states ${\cal S}_{N-i}, 0 \leq i \leq 3$,
where ${\cal S}_{N-2}$ designates a successful rebuild
on the spare disk and ${\cal S}_{N-3}$ is the failed state.
Given $v_N = v_{N-1} =1$ and the transition probabilities:
${\cal S}_{N-1} \rightarrow {\cal S}_{N-2}$: $p_s = \mu / (\mu + (N-1) \delta )$ and
we have $v_{N-2} = p$, so that the MTDDL is:
$\mbox{MTTDL}_1= 1 / (N \delta) + 1 / ((N-1) \delta)  + p/((N-2) \delta) \approx (2+p)(N \delta)$.
The MTTDL with $k$ spares is:

\vspace{-5mm}
\begin{eqnarray}
MTTDL_k = \frac{1}{N \delta} + \sum_{i=0}^k \frac{p^i}{(N-1-i)\delta}.
\end{eqnarray}
The contribution to MTTDL decreases with increasing $k$ since $p<1$.
The depletion of spare disks is not a problem 
if additional spare disks are ordered as the supply is depleted 
(see Figure 5.22 in \cite{Gibs92}).

Given a RAID1/0 array with $M$ disk pairs an approximate way to estimate MTTDL of a RAID1/0 array  
is to treat the disk pair as a single unit with an exponential failure rate,
so that $R_{array} = e^{ M / MTTDL_{pair}}$, which uses Eq.~\ref{eq:MTTDLR5} with $N=2$. 

The direct path approximation method developed in \cite{IlVe15},
which can be used to estimate the MTTDL in some interesting cases, such as RAID5/1:

\vspace{-5mm}  
\begin{eqnarray}\nonumber
\mbox{MTDDL}_{R5/1} \approx 
\frac{\mu^3}{3M(M-1)\lambda^4} = 
\frac { MTTF^4 }{ 3M(M-1) \mbox{MTTR}^3}
\end{eqnarray}
The method is applicable to Weibull and Gamma distributions and has been extended to multiple shortest paths.
Using Eq.~\ref{eq:MTTDLR5} a hierarchical reliability modeling approach 
was used in \cite{XMS+03} to get good estimate for RAID5/1 MTTDL.
\vspace{-2mm}
\begin{eqnarray}
\mbox{MTTDL}_{R1} = \frac{\mbox{MTTDL}_{R5}^2 }{  2 \mbox{MTTR}},    \hspace{2mm}
\mbox{MTTDL}_{R5} = \frac{\mbox{MTTF}_{disk}^2 }{ M(M-1) \mbox{MTTR}}, \hspace{2mm}
\mbox{MTTDL}_{R5/1} \approx \frac{ \mbox{MTTF}^4 }{ 4 M(M-1) \mbox{MTTR}^3 }.
\end{eqnarray}

\subsection*{Multivel RAID}

The approximate reliability of an RAID5/1 and RAID1/5 array can be determined 
by substituting the expression for RAID5 into the reliability expression for RAID1, and vice-versa. 
It follows that RAID5/1 is more reliable than RAID5/1.
\vspace{-3mm} 
\begin{eqnarray}
R_{R1}= 2R_{R5} - R_{R5}^2, \hspace{5mm} R_{R5}=r^M + Mr^{M-1}(1-r) \hspace{5mm} R_{5/1} \approx 1- M^3 \epsilon^3.
\end{eqnarray}
\vspace{-7mm} 
\begin{eqnarray}
R_{R5}= R^M_{R1} + M R^{M-1}_{R1}(1- R_R1), \hspace{5mm} R_{R1}= 1 - (1-r)^2 \hspace{5mm} R_{1/5} \approx 1- M(M-1) \epsilon^4.
\end{eqnarray}

\vspace{-4mm}
\subsection{Taking into Account Controller Failures}\label{sec:controller}

Three mirrored disk configurations: ${\cal C}_i, 1 \leq i \leq 3$ are considered in \cite{NgSW87}.
(a) ${\cal C}_1$ has a single controller for the duplexed disks.
(b) ${\cal C}_2$ has one controller per disk and incoming reads are routed to both disks.
(c) ${\cal C}_3$ has two controllers crossconnected to disk.
The MTTDL is determined by a CTMC with three states:
$\boldsymbol{{\cal S}}_0$: no failed disks..
$\boldsymbol{{\cal S}}_1$: single failed disk.
$\boldsymbol{{\cal S}}_2$: Both disks or the control unit has failed.
For example, given that disk and controller failure rates are $\lambda_c$ and $\lambda_d$ 
and the disk and controller repair rates are $\mu_d$ and $\mu_c$.
The transition rate for ${\cal C}_1$ are: 
$\boldsymbol{{\cal S}}_0 \rightarrow \boldsymbol{{\cal S}}_1$ $2\lambda_c$,
$\boldsymbol{{\cal S}}_1 \rightarrow \boldsymbol{{\cal S}}_0$ $\mu_d$,
$\boldsymbol{{\cal S}}_2 \rightarrow \boldsymbol{{\cal S}}_2$ $\lambda_c + \lambda _d$, and 
$\boldsymbol{{\cal S}_0 \rightarrow  \boldsymbol{\cal S}}_2$ $\lambda_d$.
These and  similar CTMCs can be solved to estimate the MTTDL and the availability,
which is the fraction of time data can be accessed.

The {\it Crosshatch Disk Array (CDA)} proposed in \cite{NgSW94} has $N \times N $ disks
which are doubly connected to horizontal and vertical busses attached to $2 N$ controllers.
Parity groups are defined over diagonal disks, so data loss occurs with 
(1) Double disk failures, 
(2) single disk plus double controller failures,
(3) quadruple controller failures.
Reliability analysis is used to show that the MTTDL attained by CDA exceeds 
the MTTDL attained with less complex organizations.


\vspace{-2mm}
\subsection{Storage Reliability Research at IBM Research at Zurich}\label{sec:Zurich}

Efficient replica maintenance and placement to attain high availability
and data durability has been the topic of research,
which lead to implementations such as Carbonite \cite{Chu+06}. 
In this section we briefly review recent research at IBM Research at Zurich. 

Replica placement in the context of large-scale data storage systems 
to attain increased availability is investigated in \cite{Ven+10}.
There are $n$ nodes and user data blocks are replicated $r$ times.
Three data placements are considered:
(a) Declustered: $r$ data replicas are placed randomly at the nodes.
(b) Clustered: nodes are divided into disjoint sets of $r$ nodes on which data is replicated.
(c) k-clustered: "the $n$ nodes  are divided  into  disjoint  sets  of $k$ nodes called clusters.
Each of these clusters is an independent storage system with $k$ nodes with a declustered  placement scheme.  
No data block in one cluster is replicated in another cluster nodes partitioned into disjoint sets of $k$ nodes
and the declustered placement is followed at these nodes".
Analytic and simulation results show that for a replication factor of two all 
placements have an MTTDL within a factor of two.

{\it Clustered Placement (CP)} and {\it Declustered Placement (DP)} is considered in \cite{VIFU11}.
DP spreads replicas across all other nodes,
while a minimum number of nodes are used by CP.
The average lifetime of a node is set to be of the order of $\delta^{-1}=10^5$ hours
and given $c$ bytes per node and a rebuild bandwidth $b$ it takes $c/b = 10$ hours to rebuild a  node, 
so that $\delta c/b \ll 1$. 
Given that the probability that the system experiences data loss is $P_{DL}$,
than $MTTDL \approx [n \delta P_{DL}]^{-1}$. 
Eq. (4) in the paper leads to $P_{DL}$. 

For $r=2$ $MTTDL^{clus.} \approx b / (n c \delta^2)$
and $MTTDL_{declus.} \approx b / (2nc \lambda^2)$,
so that the MTTDL is inversely proportional to the number of nodes.

For $r=3$ $MTTDL^{clus} \approx b^2 / ( n c^2 \delta^3)$ 
$MTTDL^{declus} \approx (n-1) b^2 / ( 4 n c^2 \delta^3)$,
which is almost independent of the number of nodes. 

Reliability of data storage systems under 
rebuild bandwidth constraints is discussed in \cite{VeIH12}.
An increased degree of replication increases parallelism for rebuild,
but this is so if sufficient bandwidth is available for this purpose.

The effect of CP and DP with two reliability metrics:
{\it Expected Annual Fraction of Data Loss (EAFDL)} and MTTDL is investigated in \cite{IlVe14}.
There are $n$ storage devices and $r=2,3$ is the replication factor.
CP (resp. DP) replicates data at the other $r − 1$ (resp. $n-1$ devices).
Rebuild time is obtained taking into account disk capacities $(c)$.
the amount of user data $U=n c /r$. 
The reserved bandwidth per device is $b$ so that time to read a disk is $c/b$.
Given that mean time to disk failure is $1/delta$ for $r=2$:

\[
MTTDL_{CP}  = b / (n c \delta^2) \mbox{  and  }  MTTDL_{DP} = b / (2n c \delta^2).
\]
\[
EAFDL_{CP} = \delta^2 c/ b \mbox{   and  } EAFDL = 2 \delta^2 c (n-1)b.
\]

A study comparing erasure coding with replication, 
in the context of peer-to-peer systems built to provide storage durability 
by taking advantage of network bandwidth, storage capacity, and computational resources is \cite{WeKu02}.
It is shown that erasure coding provides an MTTF many orders of magnitude higher than replicated systems,
while utilizing an order of magnitude less bandwidth and storage.

\vspace{-2mm}
\subsection{Simulation for Reliability Modeling}\label{sec:sim}

We have discussed the reliability analysis of small disk arrays, but more complex methods, 
possibly based on hierarchical reliability modeling, 
might be required to analyze large storage systems.
When the system is represented by a CTMC, numerical methods can be applied to solve it, 
but there is an exponential increase in the state-space,
which limits the applicability of numerical methods.

Fast simulation with importance sampling is a powerful tool applicable in this case.
\footnote{\url{https://en.wikipedia.org/wiki/Importance_sampling}.}
can be applied to reliability analysis of systems with a large number of components,

Survey of approaches for high availability solutions yielded the following statistics \cite{Endo16}: 
Experimentation 28, Simulation 3, Quantitative 1, Modeling 0.
There are few publications modeling large-scale systems such as storage clouds
and interestingly simulation has played a small role.


Hierarchical RAID (HRAID) extends the RAID paradigm to two levels \cite{ThTH12}.
There are $N$ {\it Storage Nodes (SNs)} with their own DAC and RAID$(4+\ell)$ array with $M$ disk.
There are $k$ check codes for internode redundancy, 
so that the overall redundancy level is $(k+\ell)/M$.
%
The pseudo-code for the simulator to obtain the HRAID$k/\ell$ MTTDL is given in \cite{ThTH12}. 
The code is simplified by postulating the exponential distribution for disks and DACs,
but can be extended to other failure distributions.
Modeling rebuild requires specifying repair time, 
which requires a detailed specification of the internode communication network,
at least the bandwidth available for large data transfers.

{\it Cloud Quality of Service Simulator for Reliability - CQSIM-R}
is a Monte-Carlo simulator for large scale storage systems,
which takes into account the placement of data \cite{Hall16}.
A more recent study along these lines is \cite{ZhHL17}.

\vspace{-5mm}
\section{Heterogeneous Disk Arrays}\label{sec:hetero}

RAID1 has been used in combination with other RAID levels,
which provide higher storage efficiency via lower redundancy levels.
Three examples of data stored in replicated form, later converted to erasure coding are as follows.

The {Self-Adaptive Disk Array (SADA)} starts as a RAID0/1 array and reconfigures itself onto 
an HDA with parity coding to prevent data loss and then RAID5 and RAID0 arrays \cite{PaSL06}.
The RAID0/1 array has four disk pairs $A_1$, $A_2$, $B_1$, $B_2$, $C_1$, $C_2$, $D_1$, $D_2$,
which hold copies of datasets $A$, $B$, $C$, and $D$.
If $B_1$ fails extra protection is provided for $B$ by setting $A_1 = A \oplus B$.
If $D_1$ fails the system sets $C_2= C \oplus D$.
If $D_2$ fails the system sets $A_1= (A \oplus B) \oplus (C \oplus D)$ and $C_2=D$.
In effect this is a RAID5 array with one parity disk.
Finally, if $B_2$ fails the system XORs $A$ from $A_2$, $C$ from $C_1$, 
and $D$ from $C_2$ with $A_1$ so that $A_1=B$ and we have an unprotected array.
Restriping is a generalization, which has a similarity to distributed sparing \cite{ThMe97}.
When applied to RAID$(4+k), k \geq 1$ it overwrites check strips.
In the case of RAID7 with P, Q, R check strips, overwriting starts with R strips 
and we have the transitions RAID7 $\rightarrow$ RAID6 $\rightarrow$ RAID5 $\rightarrow$ RAID0 \cite{ThTH12}   

HP's AutoRAID combines RAID1 and RAID5 arrays by initially storing data in RAID1 format,
which is converted to RAID5 format as RAID1 storage capacity is getting exhausted \cite{WGSS96}.
Hot mirroring in \cite{MoKi96} partitions storage space to areas dedicated to RAID1 and RAID5.
The main criterion to invoke migration from RAID1 to RAID5 is the number of free blocks for RAID1.
Mirrored data is placed in separate parity groups 
to ensure that they are not both affected by a disk failure in the same parity group.

DiskReduce \cite{Fan+09} initially writes three copies of data 
to accommodate the {\it Hadoop Data File Systems (HDFS)},
but then converts triplicated data to RAID5 or RAID6 formats.

{\it Heterogeneous Disk Arrays (HDAs)} combine RAID1 and RAID%
and generally parity and erasure coding in one disk array \cite{ThXu16},
whereas \cite{ThBH05} takes into account heterogeneous disks as well.
Data is allocated at the level of {\it Virtual Arrays (VAs)}, which are either RAID1 or RAID5.
Based on the volume of data and access rate a VA is allocated 
as multiple {\it Virtual Disks (VDs)} on different disks.
The VA RAID level is determined by its size and data access intensity,
e.g., VAs with high access rates to read and write small data blocks are assigned as RAID1,
while VAs with RAID5 can process large reads and writes more efficiently \cite{ThXu11}.

An advantage of HDA is that rebuild can be carried out at the level of VAs,
which is especially advantageous if they reside on disjoint drives.
Priority should be given to more critical VAs. 
This is important because the time to read multiterabyte (TB) disks 
makes them vulnerable to a second disk failure,
e.g., according to \cite{Huaw14} an 8D+1P RAID5 with 7200 RPM disks takes 40 hours for 4 TB disks.
The Huawei OceanStor RAID2.0 \cite{Huaw14} similarly to HDA speeds up 
rebuild processing by bypassing empty disk space
and XORs chunks in the same parity group to reconstruct data (see figure on page 11).

A RAID1 array combining HDD and MEMS storage is described in \cite{UyMA03}.
Even though MEMS storage costs ten times more than magnetic disks, 
hybrid MEMS/disk arrays substantially improve the performance and 
cost/performance over conventional mirrored disks.

\vspace{-5mm}
\section{Conclusions}\label{sec:conclusions}

RAID1 arrays have a 50\% redundancy in disk capacity,
which is higher then the $100k/N\%$ for RAID$(4+k)$ array with $N$ disks.
but this redundancy is acceptable in view of increased disk capacities.
RAID1 has two advantages over RAID$(4+k)$ arrays:
(1) Updating small data blocks requires writing the modified block to two disks.
while updates in RAID$(4+k)$ involves the SWP,
where each writes requires $2(k+1)$ reads and writes to update data and check blocks.
(2) RAID1 doubles disk access bandwidth to small data blocks.
On the negative side RAID$(4+k)$ allows parallel reading of strips when accessing large chunks of data
and efficient full stripe writes. 


Distributed storage systems utilize replication to ensure data availability and durability.
It is difficult to tell the difference between transient network failures and disk failures 
in wide-area bandwidth-limited systems.
To ensure an acceptable degree of replication copying is invoked 
to return the system to a desirable degree of replication \cite{Chu+06}.
A limited form of replication is remote backup database systems, 
which track a primary system and take over its operation when it fails.

\vspace{3mm} {\large \bf Commonly Used Abbreviations:} 
{\bf CD} - Chained Declustering. 
%
%
{\bf DAC} - Disk Array Controller. 
{\bf GRD} - Group Rotate Declustering. 
%
%
{\bf HDD} - Hard Disk Drive. 
{\bf ID} - Interleaved Declustering. 
{\bf LBA} - Logical Block Address. 
{\bf LSE} - Latent Sector Error. 
{\bf $k$DFT} - $k$ Disk Failure Tolerant.
{\bf MDS} - Maximum Distance  Separable.
%
%
{\bf MTTDL} - Mean Time To Data Loss. 
{\bf MTTF} - Mean Time to Failure. 
{\bf NVRAM} - NonVolatile Random Access Memory. 
{\bf OLTP} - OnLine Transaction Processing.
{\bf PCM} - Permanent Customer Model. 
{\bf RAID} - Redundant Array of Independent Disks. 
{\bf RMW} - Read-Modify-Write.
{\bf SSD} - Solid State Disk. 
{\bf SWP} - Small Write Penalty. 
%
%
{\bf VSM} - Vacationing Server Model.

\vspace{-5mm}
\section*{Appendix I: Magnetic Hard Disk Drives}\label{sec:appendix}

HDDs consist of circular disk platters coated 
with ferromagnetic material rotated at high speed, 
referred to as "spinning rust".
Data is recorded on concentric tracks formatted as 512 byte sectors,
where each sector has a header 
and a 40 byte long ECC (Error Correcting Code) \cite{JaNW08}.
Longer 4096 byte sectors, known as Advanced Format, 
provide 5-13\% higher capacity and improved error correction capability.
\footnote{\url{https://en.wikipedia.org/wiki/Disk_sector}.}

Disk capacity is increased by recording on both sides of a platter and stacking platters, 
which are served by a single disk arm with one R/W head per surface.
Tracks with the same diameter constitute a cylinder.
Disk data is modeled as a one dimensional array whose sectors are specified as 
{\it Logical Block Addresses - LBAs}, 
which are translated by the disk controller to determine LBA's location: cylinder, surface, sector number, 
which is accessed by mechanically moving the disk arm to an appropriate track,
while the disk rotation places appropriate sectors under the R/W head.
Media failures are handled by relocating bad sectors to a good area on disk:
{\it slipping} moves a faulty sector to the next physical sector as data is being written.
while {\it sparing} assigns empty sectors.

Zoned bit recording stores a higher number of sectors on its outer tracks 
to take advantage of increased circumference.
In the case of the IBM 18ES disk drives, for example, 
the number of sectors on the outermost track is approximately 60\% higher than the innermost track.
\footnote{\url{http://www.pdl.cmu.edu/DiskSim/diskspecs.shtml}.}
With a linear increase in disk sectors there would be a 30\% increase in disk capacity 
with respect to a non-zoned disk with the same number of sectors at all disks.

Disk recording density has been increasing at variable rates: 
29\%, 60\%, 100\%, 30\% per year till 1988, 1988-1996, 1997-2003, 
and since then respectively \cite{HePa07}.
The increased linear recording density combined with higher disk RPMs 
has resulted in an increased disk transfer rate (50\% in the year 2000~\cite{GrSh00}). 
{\it Shingled Magnetic Recording (SMR)},
which allows in higher areal recording densities in HDDs by writing overlapping sectors \cite{FeGi13}.

Disk access time has three components \cite{JaNW08}:
(1) Seek time to move the R/W head to the appropriate track,
(2) Rotational latency is the time for the appropriate disk block 
to rotate underneath the R/W head to be read or written 
(3) Transfer time is the time for the disk block to rotate past the R/W head.

Given the seek time characteristic of a disk $t_{seek} (d), 1 \leq d \leq C-1$, 
where $C$ is the number of disk cylinders,
the seek time is determined by the number of disk tracks traversed ($d$) by the R/W head.
For an HDD considered a recent study: 
$t_{seek}= \mbox{constant}, \mbox{ for }d \leq 100$, followed by a discontinuity,
$t_{seek} (d) \propto \sqrt{d} \mbox{ for  } 100 \leq d \leq (C-1)/3$, after which it is linear.  
Seek time decreases with increased radial density,
but there is an increase in head settling time for writes \cite{JaNW08}.
Transfer time for small disk blocks is negligible, 
so that disk access time is determined by positioning time,
which was improving at a rate of 8\% annually at the turn of the century~\cite{GrSh00}.

The rotational latency for small randomly placed disk blocks 
is one half of disk rotation time and improves with the RPM.
For 7200 RPM disks $T_{rot} = 8.33$ ms. 
so that the mean latency is $T_{rot}/2$.
Disk transfer time is negligibly short, 
e.g., for a 7200 RPM disk with 400 sectors the transfer time for a 4 KB is: 
$(8/400)\times 8.33 < 0.166$ ms.  

In view of rapidly increasing disk capacities,
disk access bandwidth to randomly placed disk blocks may be a limiting factor.
Increasing cache sizes at different levels of memory hierarchy lower the miss rate per Gigabyte (GB), 
so that the disk access rate does not increase in proportion with disk capacity.

Rather than serving disk requests in FCFS order,
disk arm scheduling is used to reduce disks access time,
so that the disk can process more request per second.
The {\it Shortest Seek Time First (SSTF)} and SCAN 
are two early methods to reduce the seek distance \cite{Thom13},
but since rotational latency is significant with respect to seek time,
the {\it Shortest Access/Positioning Time First (SATF/SPTF)} policy,
which is implemented locally at the disk controller minimizes the sum of seek time and rotational latency.
SATF outperforms the SSTF and SCAN methods at heavier loads \cite{WoGP94,Thom06b}.  

Proximal I/O described in \cite{ScSS11} is a technique to improving random disk I/O performance  
by aggregating random updates in a flash cache whose size is 1\% of disk space 
and this allows 5.3 user writes to be destaged per revolution.
Compared to update-in-place or write-anywhere file systems, 
proximal I/O provides a nearly seven-fold improvement in random I/O performance,
while maintaining near sequential data layout.
Despite the higher cost of flash memory, the overall system cost is one third of that of a 
system achieving an equivalent number of random I/O operations.

Disk performance can be improved by reducing seek distances by rearranging its data blocks 
in the middle disk cylinders resulting in the organ-pipe organization \cite{JaNW08}.
The design and implementation of a disk subsystem that adaptively reorganizes data 
is described and the resulting performance improvement verified via experiments in \cite{VoCa90}.
In a followup study frequently accessed data is shuffled and 
placed at the center of the disk surface \cite{RuWi91}.
It is noted Shuffling can backfire by placing the data of a split file at opposite ends of the organ-pipe. 
There is the issue of frequency of data shuffling and the unit to be shuffled.

A technique to reducing disk seek times is estimate frequencies by monitoring referenced blocks 
and moving frequently accessed blocks from their original locations 
to reserved space near the middle of the disk \cite{AkSa95}. 
The scheme was implemented by modifying a UNIX device driver and 
trace-driven simulations showed that seek times can be cut substantially by copying a small number of blocks.

The {\it Automatic Locality Improving Storage (ALIS)} scheme
automatically reorganizes disk blocks accessed together to place them so they can
be accessed efficiently via a sequential access \cite{HsSY05,JaNW08}.

A multimegabyte track buffer is usually provided with modern disks, 
which is mainly used for prefetching blocks associated with sequential data transfers \cite{JaNW08}.

HDDs have been classified into {\it Enterprise Storage - ES} and {\it Personal Storage - PS} \cite{AnDR03}.
ES drives used by high performance servers utilize more advanced technology 
in the form of faster seek times and higher RPMs (rotations per minute).
Disk power consumption which increases with the cube of RPM requires smaller diameter disks: 
2.5", 3.3", and 3.7" for 15,000, 10,000, and 7200 RPM drives. 
PS drives are used several hours a day, while ES drives are powered up all the time,
but disks may be powered down in archival storage.
HDDs are also classified by their interface:
ES drives use {\em Small Computer System Interface -SCSI} or {\em Fibre Channel - FC} 
and PS drives use {\em Advanced Technology Attachment - ATA} \cite{AnDR03}.
{\it Serial ATA - SATA} HDDs start wearing out after three years.

\vspace{-5mm}
\section*{Appendix II: Erasure Coded RAID Arrays}\label{sec:RAID}

RAID5 dedicates a single strip per stripe to parity,
which is the {\it eXclusive-OR - XOR} of the data strips in that stripe.
The updating of small data blocks incurs the {\it Small Write Penalty - SWP}:
two reads to access old data ($d_{old}$) and parity ($p_{old}$) blocks, 
if they are not cached, to compute $p_{new} = d_{old} \oplus d_{new} \oplus d_{old}$ and 
two writes for $d_{new}$ and $p_{new}$.
Updating small data blocks in RAID1 requires both copies of the data to be updated.

To balance disk loads for updating parity strips in RAID5 
they are placed in right-to-left repeating diagonals, 
called the left-symmetric organization \cite{Che+94}.
In an all-stripe write when all data strips are updated the check strips can be computed on the fly.
When a majority of strips in a stripe are updated the {\it ReConstruct Write (RCW)} method 
can be used by reading the remaining strips in a stripe to compute the parity without incurring RMWs \cite{Thom05a}.
These concepts are applicable to HDAs discussed in Section~\ref{sec:org}.

Requested blocks on a failed disk are reconstructed 
on demand by XORing corresponding blocks from surviving disks.
This results in the doubling of disk load due to read requests,
but this load can be reduced by adopting the {\it Clustered RAID (CRAID)} paradigm,
which uses a smaller parity group size ($G$) than the number of disks: $G<N$, 
in which case the load increase is $\alpha = (G-1) / (N-1) < 1$ \cite{MuLu90}.
Implementations of CRAID are discussed in \cite{ThBl09}.
CRAID with $G=2$ has the same level of redundancy as RAID1.

RAID5 with a single failed disks is susceptible to data loss if a second disk fails.
If a spare disk is available the rebuild process should be initiated as soon as possible
to reconstruct the data on the failed disk on the spare to return the system to its normal mode \cite{MuLu90}.
RAID5 rebuild is accomplished by reading successive 
{\it Rebuild Units (RUs)} to be XORed for reconstruction \cite{ThMe94}.
CRAID accelerates the rebuild process, 
but the writing of rebuilt blocks on the spare disk may constitute a bottleneck.
RAID5 rebuild may fail due to {\it Latent Sector Errors (LSEs)} 
and this has led to RAID6 and generally RAID$(4+k)$ $k \geq 1$ arrays, 
which are {\it $k$-Disk-Failure-Tolerant ($k$DFT)} \cite{ThBl09}.

{\it Reed-Solomon (RS)} codes can be used to protect against more than one disk failure \cite{ThBl09}.
RS codes are {\it Maximum Distance Separable (MDS)} with minimal redundancy,
i.e., it takes the capacity of $k$ disks to tolerate $k$ disk failures.
Because of the high computational cost of RS codes, parity based coding methods, 
such as EVENODD have been developed to tolerate two or more disk failures \cite{Bla+01,ThBl09} 

\vspace{-5mm}

\end{document}